\documentclass[graybox, nosecnum, natbib]{svmult}

\usepackage{mathptmx}       
\usepackage{helvet}         
\usepackage{courier}        
\usepackage{type1cm}        
\usepackage{makeidx}         
\usepackage{graphicx}        
\usepackage{multicol}        
\usepackage[bottom]{footmisc}
\usepackage{soul}            

\usepackage{aps-bibstyle}
\usepackage{bbm}
\usepackage{mathtools}
\usepackage{amsmath}
\usepackage{amssymb}

\usepackage{booktabs} 

\makeindex             

\begin{document}
\title*{Isoscalar Giant Resonances: Experimental Studies}
\author{Umesh Garg}
\institute{Umesh Garg \at Physics Department, University of Notre Dame, Notre Dame, IN 46556, USA \\  \email{garg@nd.edu}}
%
%
\maketitle
\abstract{Giant resonances -- highly collective, high-frequency oscillations of the atomic nucleus -- are the focus of this chapter. We discuss the isoscalar excitations, where the protons and neutrons oscillate in phase, up to angular-momentum transfers $L$ = 3. The procedures of experiments and data analysis employed in extracting the strength distributions associated with these resonances are discussed. The experimentally extracted strength distributions are presented, along with information on the properties of these resonances available to date. The effect of deformation of the nuclear ground state on the resonance strength distributions is discussed. Furthermore, the exciting opportunities being opened with the current and future availability of rare isotope beams the world over are expounded.}

\section{\textit{Introduction}}
Giant resonances are often considered as highly collective modes of nuclear excitation in which an appreciable fraction of the nucleons of a nucleus move
together. Indeed, the motion is so collective that it is quite appropriate to think of these modes of excitation in hydrodynamic terms like the oscillation of a liquid
drop. Microscopically, the giant resonances are interpreted as coherent superpositions of all possible 1p-1h transitions between the states of the shell model induced by a single-particle operator. In this simple picture, they should appear as even- or odd-parity multipole groups separated by the gap of the shells (1$\hslash\omega \sim$ 40 A$^{-1/3}$ MeV). The residual particle-hole interaction tends to concentrate the transition strength into a few collective states.  
If protons and neutrons oscillate in phase, the excitation is isoscalar ($T$ = 0), and isovector ($T$ = 1) if out of phase; likewise spin-up and spin-down nucleons moving in phase lead to electric ($S$ = 0) modes while the movement out-of-phase gives the so-called magnetic ($S$ = 1) modes. In principle, all modes of angular-momentum transfer ($L$) are feasible, although not much experimental information has been available for modes with $L \geq$ 3.

The interaction for inelastic scattering can excite a nucleon by at most $L \hslash\omega$ or, in other words, the nucleon can be promoted by at most $L$ major shells. The number of shells is either odd or even according to the parity.
Thus, the $L$ = 2 transitions can correspond to 0$\hslash\omega$ (transitions within the same shell; the 2$^+$ states in the even-even nuclei, for example) or 
2$\hslash\omega$. The latter excitation is the isoscalar giant quadrupole resonance (ISGQR). Likewise, $L$ = 3 would have transitions with 1$\hslash\omega$ (the low-energy octupole resonance: ISLEOR) and 3$\hslash\omega$ (the high-energy octupole resonance: ISHEOR). The $L$ = 0 (the isoscalar giant monopole resonance: ISGMR) and $L$ = 1 (the isoscalar giant dipole resonance: ISGDR) modes are special in that they correspond to 2$\hslash\omega$ and 3$\hslash\omega$  transitions [see the discussion in the Chapter by G. Col\`{o}].

There are some properties common to all giant resonances:
\vspace*{-2mm}
\begin{itemize}
\item they are a general property of nuclei, present in every nucleus;
\item the excitation energy of each resonance has a roughly $A^{-1/3}$ dependence;
\item they exhaust a large fraction of the so-called energy-weighted sum rule (EWSR) for the particular mode.
\end{itemize}
\vspace*{-2mm}
The ``sum-rule'' refers to the {\it total} transition strength allowable quantum mechanically for a mode with specific ($L, T, S$) values. The EWSR is preferred because it is nearly model-independent.

The history of giant resonances goes back to the late 1940's, when what is now known as the isovector giant dipole resonance (IVGDR: $L$ = 1, $T$ = 1, $S$ = 0) was observed as a large peak in the photofission cross sections when U and Th nuclei were bombarded with $\gamma$-rays produced with a 100 MeV betatron (\cite{IVGDR1}). For the next two decades or so ``giant resonance'' meant the IVGDR. In early 1970's, the ISGQR was identified in inelastic scattering of 65-MeV electrons (\cite{Pithan1971}) and 62-MeV protons (\cite{Bertrand1972}). The next resonance to be experimentally observed was the ISGMR in the mid-to-late 1970's, in ``small angle'' inelastic scattering of 82-MeV deuterons (\cite{ddprime1}) and 96-120 MeV $\alpha$ particles (\cite{Harakeh_prl1977,dhy1}). Since then, detailed investigations of giant resonances have been carried out using different probes. It may be said that it is now a ``mature'' field of research, which saw significant activity and enthusiasm for more than four decades beginning in the 1970's, but still remains of interest to nuclear physicists and, increasingly, to astrophysicists. Along the way, there have been several review articles recording these developments [for example, (\cite{bertrand1981,vWd1})], as also a monograph (\cite{Harakeh_book}) that is, for all practical purposes, the ``textbook'' of this field. A review of the ISGMR and ISGDR has appeared very recently (\cite{garg-colo2018}); these so-called ``compressional'' modes have been of particular interest to the wider physics community because they provide an ``experimental'' value for  the nuclear incompressibility, $K_\infty$, one of the three parameters characterizing the equation of state of infinite nuclear matter at saturation density, with crucial implications on the studies of myriad astrophysical phenomena (neutron stars, for example).

This Chapter will concern itself with experimental investigations of the isoscalar electric giant resonances up to $L$ = 3,  basically beyond the monograph by Harakeh and van der Woude, on which it draws heavily, along with (\cite{garg-colo2018}), from which some parts have been adopted nearly verbatim. Some other giant and pygmy resonances are covered in chapters by Zegers, and Zilges and Savran. 

\section{\textit{Collection and Analysis of Experimental Data}}
Experimental determination of the giant resonance strength distributions has been accomplished generally via inelastic scattering of isoscalar particles---typically $\alpha$ particles or, in some cases, deuterons---at energies of 35-100 MeV/nucleon. Other probes utilized in these studies have been electrons (60-300 MeV), protons (60-800 MeV), $^3$He (108-120 MeV), and $^6$Li (156-600 MeV). Here, we will restrict ourselves primarily to results from inelastic $\alpha$ scattering, which has proven to be the most effective tool for detailed investigation of the isoscalar giant resonances over the years; indeed, it would not be an exaggeration to call the $\alpha$-particle the workhorse of giant-resonance studies. In particular, we will present results from measurements carried out at the Research Center for Nuclear Physics (RCNP), Osaka University, Japan, using ``small angle'' inelastic scattering of 386-MeV $\alpha$'s. These measurements have provided reliable strength distributions for ISGMR, ISGDR, ISGQR and ISHEOR
in a large number of nuclei over the mass range A = 16--208, especially emphasizing the work on the ISGMR because of its importance to understanding the nuclear equation of state and the myriad astrophysical phenomena. 

\begin{figure}[!h]
\vspace{-0.3in}
\centering\includegraphics [height=0.38\textheight]{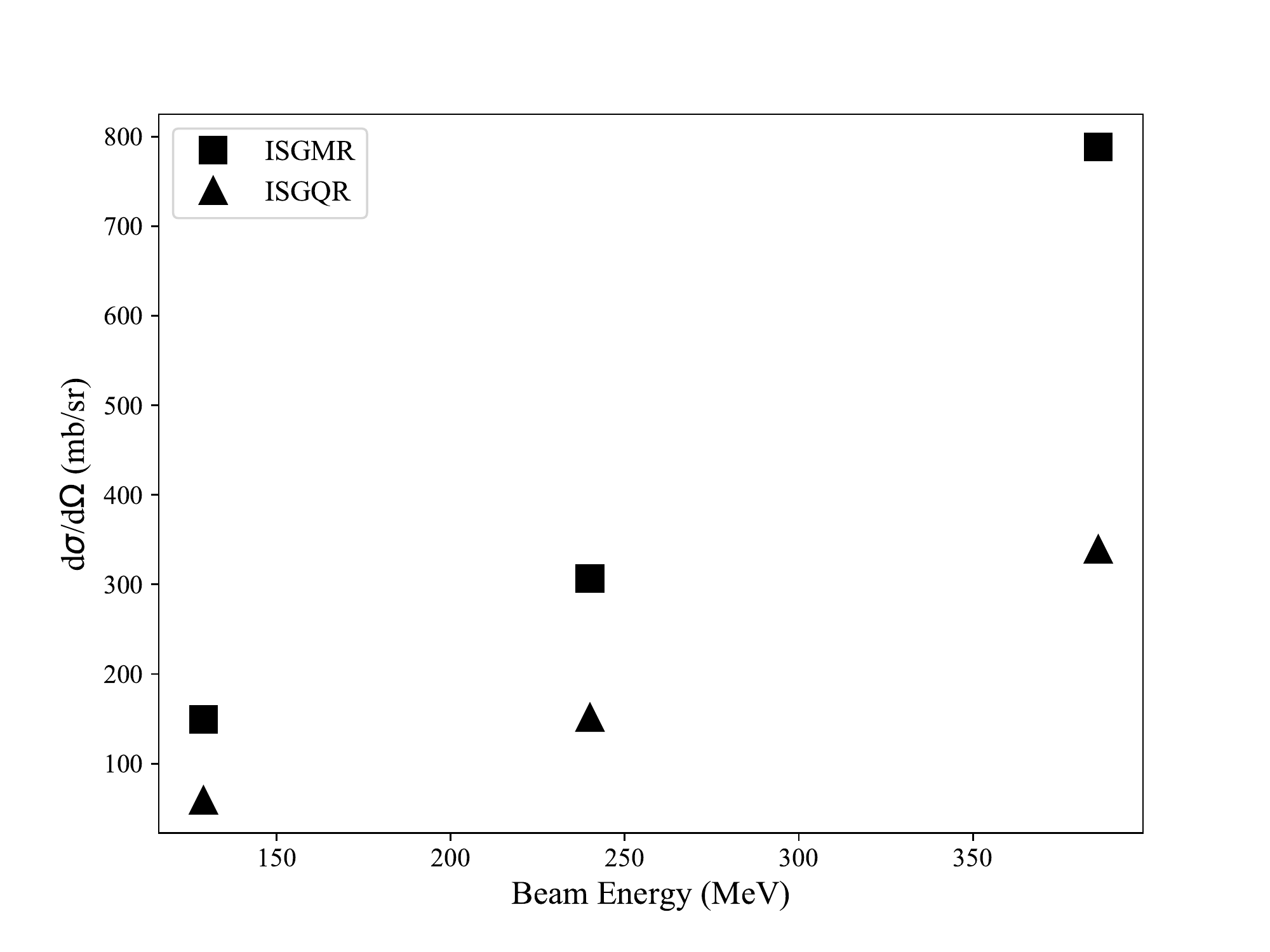}
\caption{
Differential cross sections from DWBA calculations for excitation of ISGMR at 0$^\circ$ (solid squares) and ISGQR at the first maximum (solid triangles) in inelastic scattering of 386-MeV $\alpha$ particles off $^{116}$Sn. Each calculation corresponds to an excitation energy $E_x$ = 15.8 MeV and exhaustion of 100\% of the EWSR in a single state at that excitation energy.}
\label{crosssections}
\end{figure}

As may be seen in Fig. \ref{crosssections}, the cross
sections for excitation of ISGMR rise sharply with incident $\alpha$ energies up to $\sim$400 MeV, after which the rise is quite modest (\cite{Bonin1984}). A similar, albeit less dramatic, increase is observed for the excitation of ISGQR (and other modes as well).

Because of the highly absorptive nature of the $\alpha$-nucleus and deuteron-nucleus interactions, the scattering may be treated as off a ``black disk'', with the cross 
sections, to the first order, given by the squares of the corresponding Bessel functions, $J_\lambda$ (\cite{Bessel}):

\begin{equation}
\left(\frac{d\sigma}{d\Omega}\right)_{0^+\rightarrow 0^+} \propto  |J_0(qR_D)|^2,
\end{equation}

\begin{equation}
\left(\frac{d\sigma}{d\Omega}\right)_{0^+\rightarrow 1^-} \propto |J_1(qR_D)|^2,
\end{equation}

\begin{equation}
\left(\frac{d\sigma}{d\Omega}\right)_{0^+\rightarrow 2^+} \propto \left[\frac{1}{4}J_0(qR_D)^2 + \frac{3}{4}J_2(qR_D)^2\right].  
\end{equation}
\noindent
Here, $q$ is the momentum transfer, and $R_D$, the diffraction radius, which is adjusted to fit the phase of the elastic scattering angular distribution. 
This leads to rather

\begin{figure}[!h]
\vspace{-0.3in}
\centering\includegraphics [height=0.35\textheight]{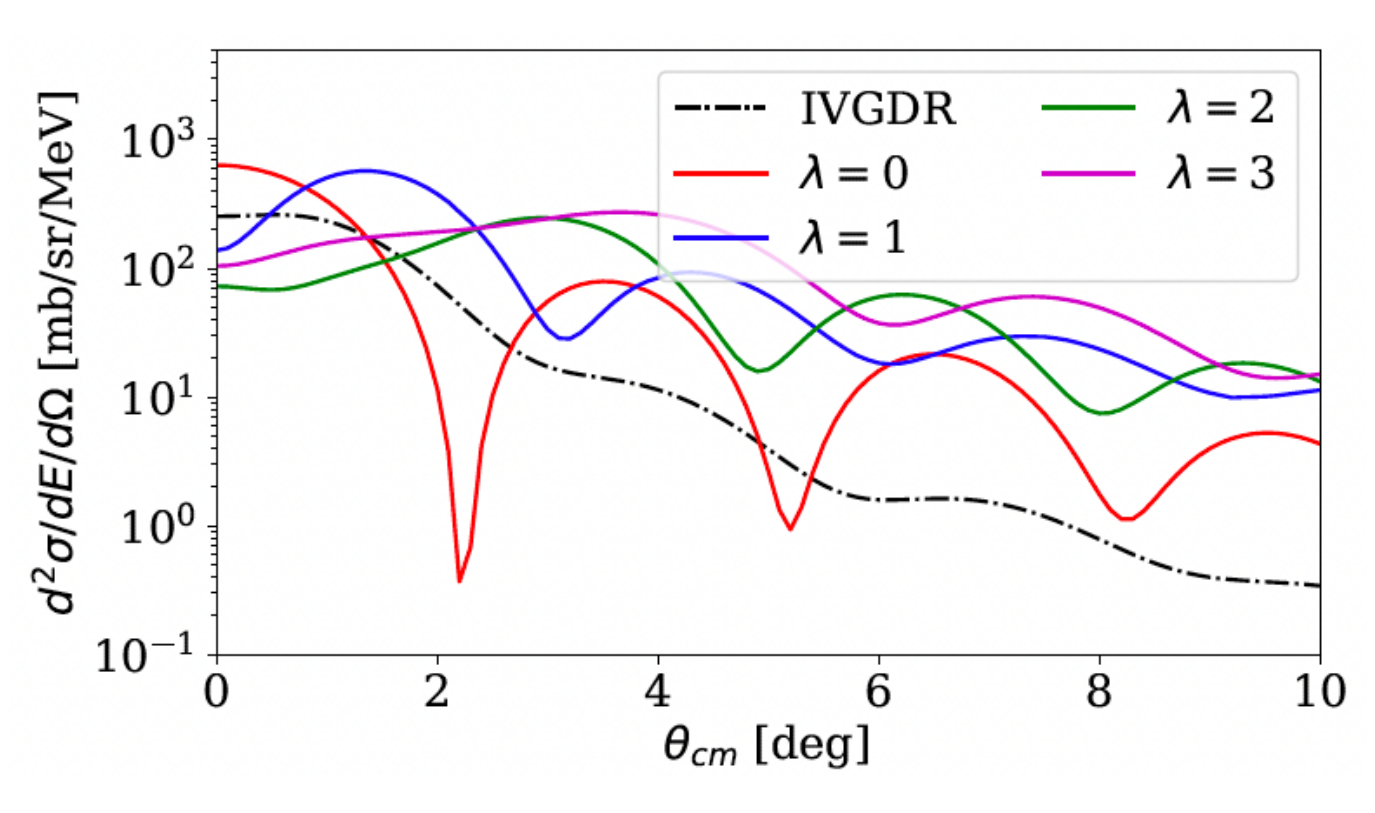}
\caption{
DWBA calculations of angular distributions of differential cross sections for excitation of isoscalar states of various multipoles ($L$ = 0--3, as well as the IVGDR) at an excitation energy of 15.0 MeV in inelastic scattering of 386-MeV $\alpha$ particles off $^{94}$Mo. Each calculation corresponds to exhaustion of 100\% of the EWSR in a single state at that energy. The letter $\lambda$ used in this figure is often substituted for $L$. Figure from (\cite{KBHoward_thesis}).}
\label{angdist}
\end{figure}

\noindent
distinctive angular distributions of the inelastic scattering cross sections for various multipoles. Fig. \ref{angdist} shows distorted-wave Born approximation (DWBA) calculations
for angular distributions of differential cross sections for inelastic scattering of 386 MeV $\alpha$ particles off $^{94}$Mo for excitation of a state at an excitation energy of 15.0 MeV corresponding to  angular momentum transfers $\Delta${\em L} = 0-3, and full exhaustion of the respective EWSR's.

\begin{figure}[h!]
\centering\includegraphics [height=0.40\textheight]{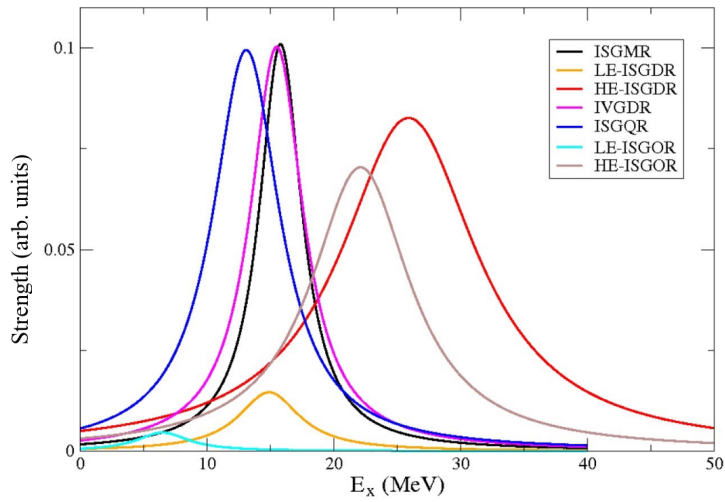}
\caption{
Hypothetical strength distributions for various electric isoscalar giant resonances (plus the IVGDR) in the nucleus $^{116}$Sn with representative energies and widths. Figure from (\cite{Patel_Thesis}).}
\label{multiple-res}
\end{figure}

A complication in this procedure arises because of the overlaping excitation energies of various giant resonances (see Fig. \ref{multiple-res}). For example, the ISGMR ($L$ = 0) overlaps significantly with the ISGQR ($L$ = 2); and the ISGDR ($L$ = 1) with the so-called high-energy
octupole resonance (ISHEOR, $L$ = 3). 

As is clear from Fig. \ref{angdist}, the angular distributions corresponding to these overlapping $L$ values are clearly distinct only at very forward angles ($\leq$ 5$^{\circ}$). At higher beam energies, the angular distributions would be further ``compressed'', making the angular range for distinctive multipolar characteristics smaller, and the measurements even more difficult to carry out successfully. 
Thus, one needs to measure inelastic scattering at center-of-mass energies of 35-100 MeV/nucleon and at very small angles to clearly identify the strengths corresponding to various multipoles. The practical requirements for such measurements, therefore, are:
\begin {itemize}
\item
An accelerator -- typically, a cyclotron -- capable of providing beams at energies of 35-100 MeV/nucleon. At these energies, the cross sections for excitation of the low-$L$ giant-resonance modes are sufficient to carry out the measurements in a reasonable time. At lower beam energies, the cross sections are rather low; at higher energies, the increase in cross sections, as mentioned above, is rather small and there are practical difficulties with determination of multipolarities based on angular distributions. It is also extremely important that the beam be free of any ``halos'' or ``wings'' at the target position; indeed, the process of obtaining a ``clean'' beam requires the right equipment and expertise, and can sometimes take a large fraction of the beam time typically allotted for such measurements.
\item
A high-quality magnetic spectrometer to allow for measurements at extremely forward angles, including 0$^{\circ}$, since only at small angles is it possible to clearly distinguish the various multipoles based on their distinct angular distributions. The 0$^{\circ}$ measurement is crucial because the ISGMR cross section is maximal there (see Fig. \ref{angdist}). 
[An exception to this requirement occurs when using inverse-kinematics reactions (with radioactive ion beams, for example); those measurements are discussed later.]
\end {itemize}

These requirements, in principle, are met at several laboratories around the world. However, almost all recent measurements on the isoscalar electric giant resonances with stable beams have been carried out at the Research Center for Nuclear Physics (RCNP), Osaka University, and at the Texas A \& M University Cyclotron Institute (TAMU) (with some decay measurements carried out at KVI, Groningen; there are also measurements at RCNP with deuterons, which are discussed separately later in this Chapter). In the RCNP measurements, a 386-MeV $\alpha$-particle beam is employed and inelastic scattering spectra are measured using the magnetic spectrometer Grand Raiden (\cite{fujiwara99}) for detection of the scattered particles. The TAMU group uses 240-MeV $\alpha$ beams and the MDM spectrometer (\cite{pringle_mdm}). The scattered particles are momentum analyzed by the spectrometers and focused onto the focal-plane detector systems comprising a combination of multi-wire drift chambers, proportional counters, ionization chambers, and scintillation detectors, to allow for particle identification and determination of position (both x- and y- in case of RCNP work) and the angle of incidence of the scattered particle via the ray-tracing technique; typically, x-position resolutions of $<$1.0 mm and angular resolution of $<$0.15$^{\circ}$ have been achieved. The 0$^{\circ}$ measurements present a special challenge because the beam itself also has to go through the system. In the TAMU arrangement, the beam passes beside the focal-plane detector and is stopped on a carbon block inside a Faraday cup behind the detector (\cite{dhy1998}), whereas in the RCNP system, it passes through a hole in the detectors and is stopped in a Faraday cup (FC) placed several meters downstream from the detectors; Fig. 1 of (\cite{Itoh_prc2003}) shows the arrangement of the three Faraday cups utilized in the RCNP measurements for different angular ranges.
In both cases, inelastic scattering spectra are measured typically over the angular range  0$^{\circ}$--10$^{\circ}$, with elastic scattering spectra also measured (over a much wider angular range, typically going from $\sim$3$^{\circ}$ to beyond 25$^\circ$) in many cases, in order to obtain appropriate optical-model parameters used in the DWBA calculations, as described hereinafter.
Calibration of the excitation-energy spectra is carried out by measuring elastic and low-energy excitation peaks in the nuclei $^{12}$C and $^{24}$Mg.

The use of magnetic spectrometers in giant-resonance studies is discussed in detail in the chapter by Fujiwara.

\begin{figure}[h!]
\centering\includegraphics [height=0.60\textheight]{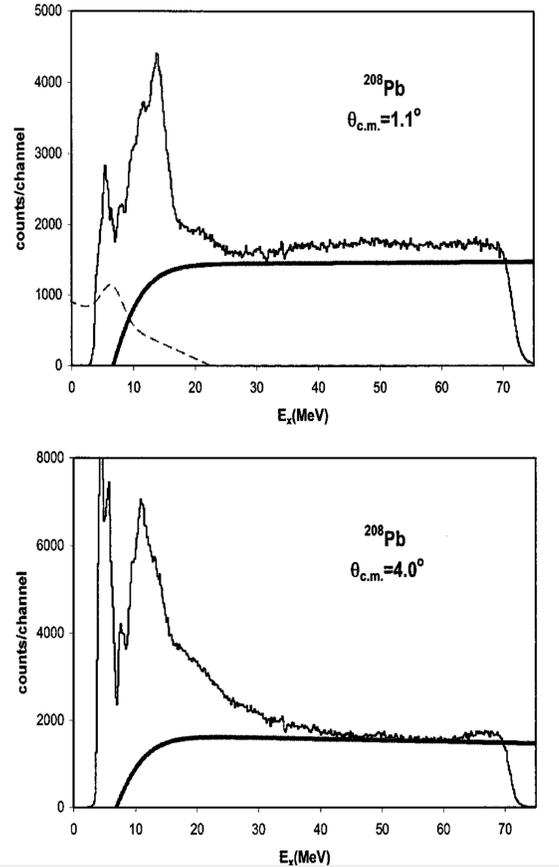}`
\caption{Excitation-energy spectra for $^{208}$Pb from the TAMU ($\alpha, \alpha'$) work at the scattering angles indicated. The thick solid lines show the continuum chosen for the analysis. The dashed line below 22 MeV represents a contaminant present at some angles in the spectra taken with the MDM spectrometer at 0$^{\circ}$. Figure from (\cite{dhybg}).}
\label{tamu_bg}
\end{figure}

The inelastic scattering spectra for the medium-mass and heavier nuclei (A$\geq$90) typically consist of one or two broad giant-resonance ``bumps'' on top of a ``background'' (see Fig. \ref{tamu_bg}). This ``background'' comprises excitation of nuclear continuum

\noindent 
and contributions from three-body channels, such as knock-out reactions (\cite{bran1}). At lower beam energies ($<$ 40 MeV/nucleon), there are also the contributions from the so-called ``pick-up and decay'' channels, whereby the incoming $\alpha$ particle picks up a proton or a neutron from the target, forming $^{5}$Li and $^{5}$He, respectively. These unstable nuclei decay almost immediately, with the final $\alpha$ particle leading to a spurious ``bump'' in the inelastic scattering spectra, not dissimilar to a giant resonance. This is purely a kinematical effect and the position of  this ``bump'' depends on the scattering angle and beam energy. Still, this effect may lead to claims of identification of new resonances, one case being that of the ISHEOR in $^{208}$Pb using $\sim$20 MeV/nucleon $^{16}$O particles (\cite{doll_prl}); later measurements, using
a $^{14}$N beam at 19 MeV/nucleon, clearly established (\cite{garg_N14}) that there was no evidence for excitation of resonances purported to have been observed in (\cite{doll_prl}). 

The largest contributions to this ``background'' are, generally, instrumental, originating from re-scattering of elastically scattered particles from the opening slits and other parts of the spectrograph. This problem is quite severe at small angles where the elastic scattering cross sections are very large. This large overall ``background'' had been a bane of all giant resonance measurements for the longest time because there is no direct way to calculate, or even estimate, its shape and magnitude. What one did was to subtract out from the spectra a background of  ``reasonable'' shape
before further analysis. The ``reasonable shape'' could be a matter of debate, of course, and the process always led to questions about the correctness of the extracted results.

Of the two aforementioned laboratories from where most of the recent giant resonance measurements have come, the TAMU group uses this background subtraction process still. After extensive investigations of the results of different background shapes, they now employ an empirical background 
assuming that it has the shape of a straight line at high excitation, joining onto a Fermi shape at low excitation to model particle threshold effects (\cite{dhybg,dhybg2}); an example of such empirical background is shown in Fig. \ref{tamu_bg}. 

\begin{figure} [h]
\centering\includegraphics [height=0.35\textheight]{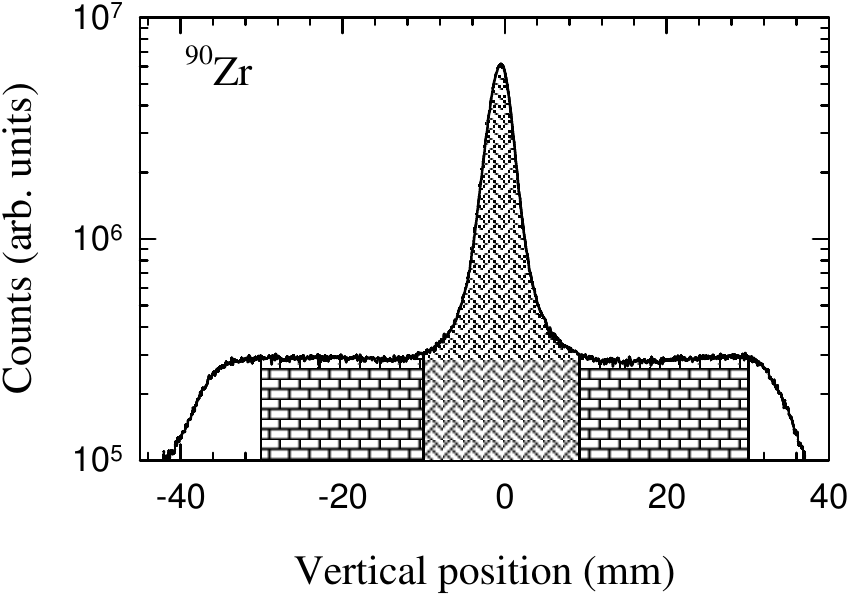}
\caption{Vertical-position spectra from the RCNP $^{90}$Zr ($\alpha, \alpha'$) work, with the Grand Raiden spectrometer set at 0$^{\circ}$. The central region represents true+background events. The off-center regions represent only background events. The true events were obtained by subtracting background events from the true+background events; ``True'' stands for ``target-scattered''. Figure from (\cite{YKG_A90_PRC}).}
\label{bgsubtraction}
\end{figure}

The spectra in the RCNP measurements, on the other hand, are essentially free of all instrumental background. 
The ion-optics of Grand Raiden enables particles scattered from the target position to be focused vertically at the focal plane. On the other hand, background events due to the re-scattering of $\alpha$ particles from the wall and pole surfaces of the spectrometer show a flat distribution in the vertical-position spectra at the focal plane, as shown in Fig. \ref{bgsubtraction} for $^{90}$Zr($\alpha,\alpha'$) with the Grand Raiden spectrometer set at 0$^{\circ}$. The vertical center densely hatched region contains a combination of true (target-scattered) events plus those from the background; the off-center sparsely hatched regions comprise only the background. 

\begin{figure} [!h]
\centering\includegraphics [height=0.60\textheight]{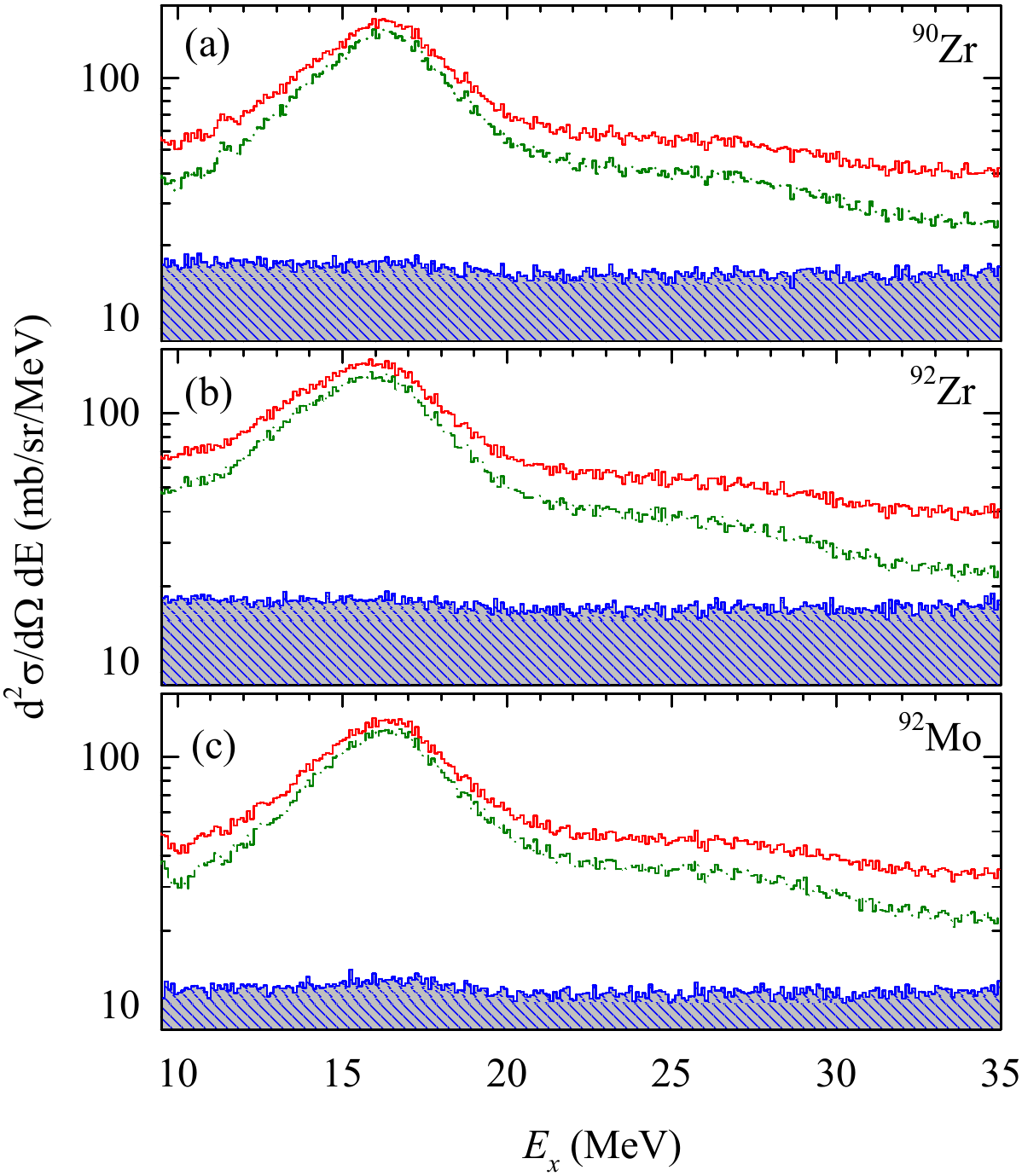}
\caption{Excitation-energy spectra for the ($\alpha, \alpha'$) reaction at $E_{\alpha}$ = 385 MeV at an average spectrometer angle of $\theta_\mathrm{avg}$ = 0.7$^{\circ}$ for $^{90}$Zr, $^{92}$Zr, and $^{92}$Mo in panels (a), (b), and (c), respectively. In each panel, the blue hatched region represents the instrumental background.
The solid red and green histograms show the energy spectra before and after the instrumental-background subtraction, respectively. Figure from (\cite{YKG_A90_PRC}).}
\label{subtracted-spectrum}
\end{figure}

\begin{figure} 
\centering\includegraphics [height=0.50\textheight]{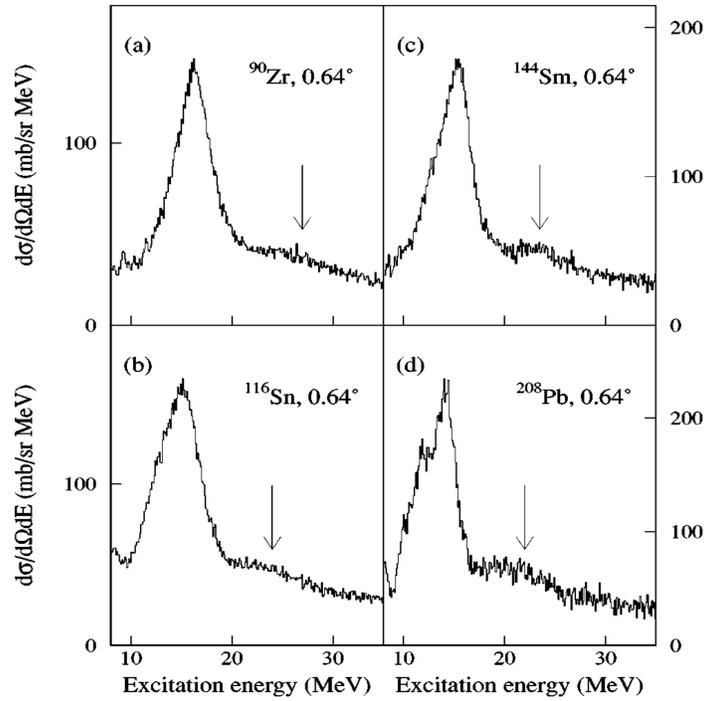}
\caption{``Background-subtracted'' excitation-energy spectra for $^{90}$Zr, $^{116}$Sn, $^{144}$Sm, and $^{208}$Pb from the RCNP ($\alpha, \alpha'$) work at an average scattering angle of $\theta_\mathrm{avg}$  =  0.64$^{\circ}$.  The main ``bump'' consists primarily of ISGMR and ISGQR; the secondary ``bump'' is ISGDR+ISHEOR. The arrows indicate the location of the ISGDR as extracted in that work. Figure from (\cite{Uchida_90Zr})}
\label{0degspectra}
\end{figure}

Figs. \ref{subtracted-spectrum} (a) to (c) show the instrumental background, and excitation-energy spectra before and after the background subtraction, as measured at an average spectrometer angle  $\theta_\mathrm{avg}$ = 0.7$^{\circ}$ for $^{90,92}$Zr and $^{92}$Mo at 0$^{\circ}$. The instrumental-background spectrum has no discernible structures in the giant resonance region. Clean ``true'' spectra are obtained by subtracting
the instrumental background spectrum from the true+background spectrum. These spectra have, practically, no instrumental contributions to the ``background'', which now comprises the true nuclear continuum and, at the highest excitation energies, contributions from three-body channels, such as knock-out reactions, as mentioned earlier in the text. Representative ``background-subtracted'' inelastic scattering spectra for several nuclei from RCNP work are shown in Fig.~\ref{0degspectra}. Incidentally, a similar procedure was used many years ago in giant resonance studies at KVI, Groningen (\cite{bran1}).

The background-subtracted spectra, so obtained, are used in a multipole-decomposition analysis (MDA) (\cite{Bonin1984, Li_2010}) to extract the multipole strength distributions. In the MDA process, the experimental cross sections at each angle are binned into small (typically, $\leq$1 MeV) excitation-energy intervals. The laboratory angular differential cross sections for each excitation-energy bin are converted to the center-of-mass frame using the standard Jacobian and relativistic kinematics. For each excitation-energy bin, the experimental angular differential cross sections, $\frac{d\sigma^{\rm exp}}{d\Omega}(\theta_{\rm c.m.},E_x)$, are fitted by
means of the least-square method with the linear combination of
the calculated double-differential cross sections associated with different multipoles:

\begin{equation}
\frac{d^{2}\sigma ^{\mathrm{exp}} (\theta_{\mathrm{c.m.}}, E_{x})}{d\Omega dE} =  \sum\limits_{L=0}a_{L}(E_{x})\frac{d^{2}\sigma_{L}^{\mathrm{DWBA}}(\theta_{\mathrm{c.m.}}, E_{x}) }{d\Omega dE},  
\label{MDA}
\end{equation}
\noindent
where $a_{L}(E_{x})$ is the EWSR fraction for the $L^{\rm th}$ component, and $\frac{d^{2}\sigma_{L}^{\mathrm{DWBA}} }{d\Omega dE} (\theta_{\mathrm{c.m.}}, E_{x})$ is the cross section corresponding to 100\% EWSR for the $L^{\rm th}$ multipole at excitation energy $E_{x}$, calculated using DWBA. The fractions of the EWSR, $a_L(E_x)$, for
various multipole components are determined by minimizing $\chi^2$.
This procedure is justified since the angular distributions are well
characterized by the transferred angular momentum $\Delta L$, according to the DWBA calculations for $\alpha$ scattering. For the limited angular range covered in these measurements, summation over $L\leq$7 is more than sufficient to extract the desired strength distributions; indeed, it is not possible to reliably extract the strength distributions for $L\geq$4 
over this limited angular range. The uncertainties in the $a_L(E_x)$ coefficients are estimated by changing the magnitude of the one component $a_{L}(E_{x})$, until refitting by varying the other components resulted in an increase in the $\chi^{2}$ by 1 (\cite{YB_24Mg_1999,Itoh_prc2003}). In more recent work, these uncertainties have been estimated using a Monte Carlo sampling from the probability distributions of the individual $a_{L}(E_{x})$, and constitute a 68\% confidence interval (\cite{KBH_CA, KBH_Mo}).

The computer codes PTOLEMY (\cite{ptolemy1, ptolemy2}) and ECIS95 (\cite{ecis}) were used to perform the DWBA calculations, with the input values in PTOLEMY modified (\cite{Satchler1992}) to take into account the correct relativistic kinematics. The shape of the real part of the potential
and the form factor for PTOLEMY were obtained using the codes
SDOLFIN and DOLFIN (\cite{dolfin}). The transition
densities and sum rules for various multipolarities employed in these calculations are obtained from 
(\cite{Harakeh_book, Satchler1987, Harakeh1981}) and the radial moments obtained by numerical integration of the Fermi mass distribution using the parameters $c$ and $a$ from (\cite{Fricke1995}).

Even though the $\alpha$ particle is isoscalar, the isovector giant dipole resonance (IVGDR) is excited at these beam energies via Coulomb excitation. The cross sections for IVGDR excitation increase with increasing beam energy and can be quite significant, especially for heavy target nuclei (\cite{izumoto, shlomo-1}). Because its energy is nearly identical to that of the ISGMR, the IVGDR contribution has to be properly accounted for in the MDA. This is carried out (\cite{Darshana2012, Li_2010}) by employing IVGDR parameters 
from previously-known photonuclear cross-section data (\cite{Berman_1975}) in conjunction with DWBA calculations based on the Goldhaber-Teller model to estimate the IVGDR differential cross sections as a function of scattering angle (\cite{Satchler1987}).

To perform the DWBA calculations, one requires appropriate optical-model parameters (OMPs). For this purpose, data are obtained for elastic scattering (and inelastic scattering to the low-lying states) over a wide angular range (typically, $\geq 3^{\circ}$--30$^{\circ}$) and the OMPs extracted from fits to the angular distributions of differential cross sections of elastic scattering. 

A ``hybrid'' optical-model potential (OMP) proposed by Satchler and Khoa (\cite{Satchler_Khoa1997}) has been employed in most of  
the RCNP and all TAMU measurements reported here. The real part of the optical potential is generated by single-folding with a density-dependent Gaussian $\alpha$-nucleon interaction (\cite{Li_2010}), and a Woods-Saxon form is used for the imaginary term. Thus, the total  $\alpha$-nucleus ground-state potential is given 
by:
\begin{equation}
U(r) = -V(r)-\it{i}W/\{1+\exp[(r-R_{I})/a_{I}]\}, 
\label{OpticalPot}
\end{equation}
where $V(r)$ is the real single-folding potential obtained using computer code SDOLFIN (\cite{dolfin})  by
folding the ground-state density with the density-dependent
$\alpha $-nucleon interaction:
\begin{equation}
\mathrm{\upsilon_{DDG}}(\mathbf{r},\mathbf{r'},\rho) = -\upsilon[1-\beta \rho(\mathbf{r'})^{2/3}]\mathrm{exp}\left(-\frac{\mathbf{|r-r'|}^2}{t^2}\right). 
\end{equation}
Here, $\mathrm{\upsilon_{DDG}}(\mathbf{r},\mathbf{r'},\rho)$ is the density-dependent $\alpha$-nucleon interaction, $\mathbf{|r-r'|}$ is the distance between center-of-mass of the $\alpha$-particle and a target nucleon, $\rho(\mathbf{r'})$ is the ground-state density of the target nucleus at a position $\mathbf{r'}$ of the 
target nucleon, $\beta$ = 1.9 fm$^{2}$, and $t$ = 1.88 fm. In Eq. (\ref{OpticalPot}), $W$ is the depth of the Woods-Saxon type imaginary part of the potential, with the radius $R_{I}$ and diffuseness $a_{I}$.

\begin{figure}[h]
\centering\includegraphics [height=0.53\textheight]{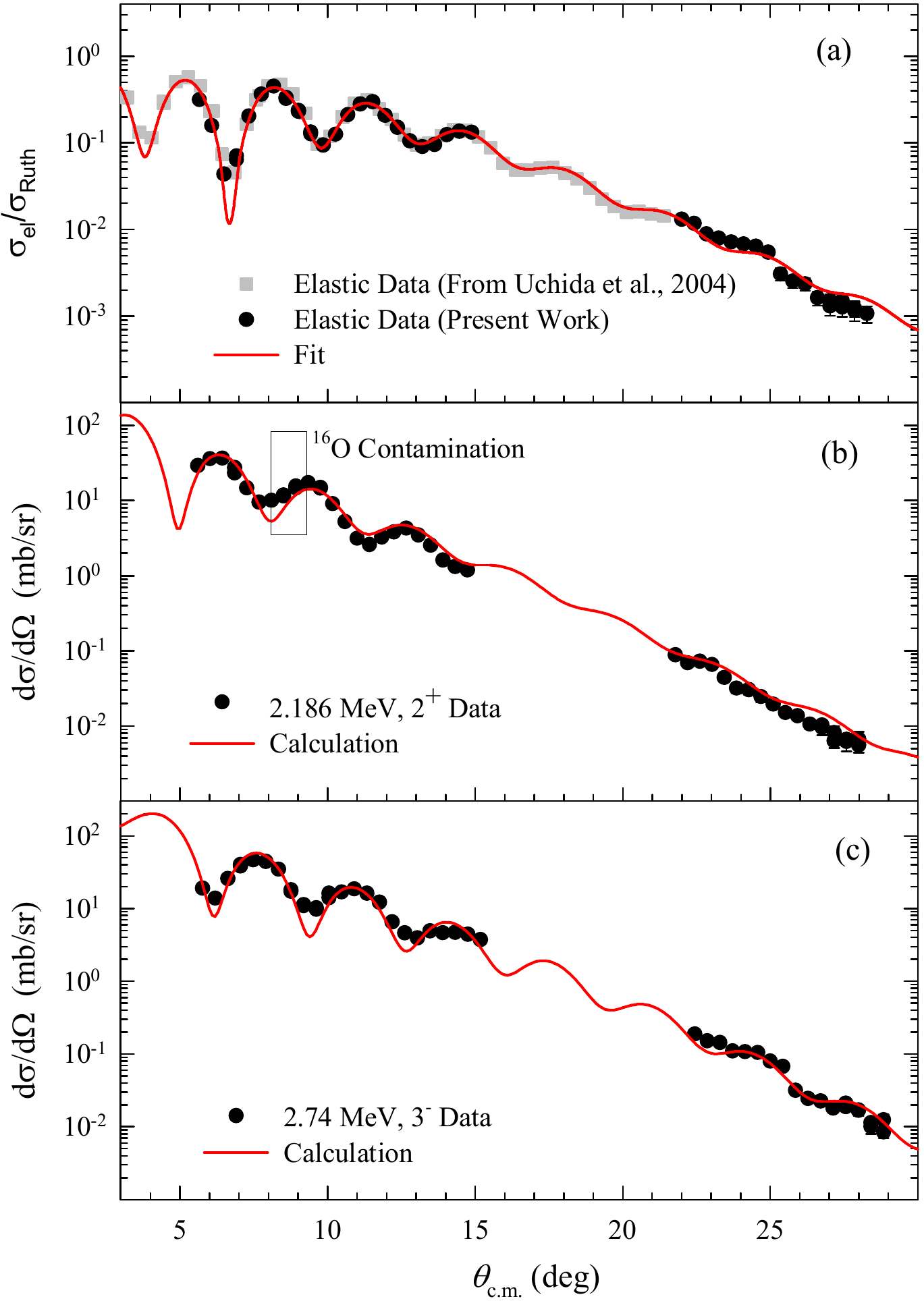}
\caption{(a) Angular distribution of the ratios of the elastic $\alpha$ scattering cross sections to the Rutherford cross sections for $^{90}$Zr at an $\alpha$ energy of 386 MeV (solid black circles). The solid red line shows a fit from the optical-model potential given in the text. (b) Differential cross sections for excitation of the 2$^+$ state in $^{90}$Zr The solid red line shows the calculated cross sections for the state using the OMPs obtained from the fits to the elastic-scattering data in (a) and the $B(E2)$ value from (\cite{BE2_24Mg}). (c) Same as (b), but for the 3$^-$ state, with the $B(E3)$ value from (\cite{BE3}). Figure from (\cite{YKG_A90_PRC}).}
\label{OMfits}
\end{figure}

The imaginary potential parameters ($W$, $R_{I}$, and $a_{I}$), together with the depth of the real part, $V$, are obtained by fitting the elastic-scattering cross sections. The appropriateness of the OMPs so obtained is tested by calculating the cross sections for the low-lying 2$^+$ and/or 3$^-$ states using these parameters and the previously-known transition probabilities for these states (\cite{BE2_24Mg, BE3}), and comparing those with the experimental values. Fig. \ref{OMfits} shows the fit to the elastic $\alpha$-scattering data from $^{90}$Zr and the comparison of the experimental and calculated differential cross sections for the first 2$^+$ and 3$^-$ states in $^{90}$Zr as an illustrative example of this procedure. Note that there is no ``fitting'' involved in Figs. \ref{OMfits}(b) and \ref{OMfits}(c) and the calculations are performed for the adopted values for the $B(E2)$ and $B(E3)$ from (\cite{BE2_24Mg, BE3}), respectively.
 
We note here that in some cases of RCNP work (\cite{Uchida_PLB2003,Uchida_90Zr,Itoh_plb,Itoh_prc2003}), the single-folding model was used, in the same form, for both the real and imaginary parts; the final strength distributions were not appreciably different from those obtained with the aforementioned ``hybrid'' model, however. A similar ``non-hybrid'' model was used also by the TAMU group in analysis of $^{6}$Li-scattering data (\cite{DHY2009_MgSi,DHY2009_Sn}).

\begin{figure*}[!h]
\centering\includegraphics [height=0.39\textheight]{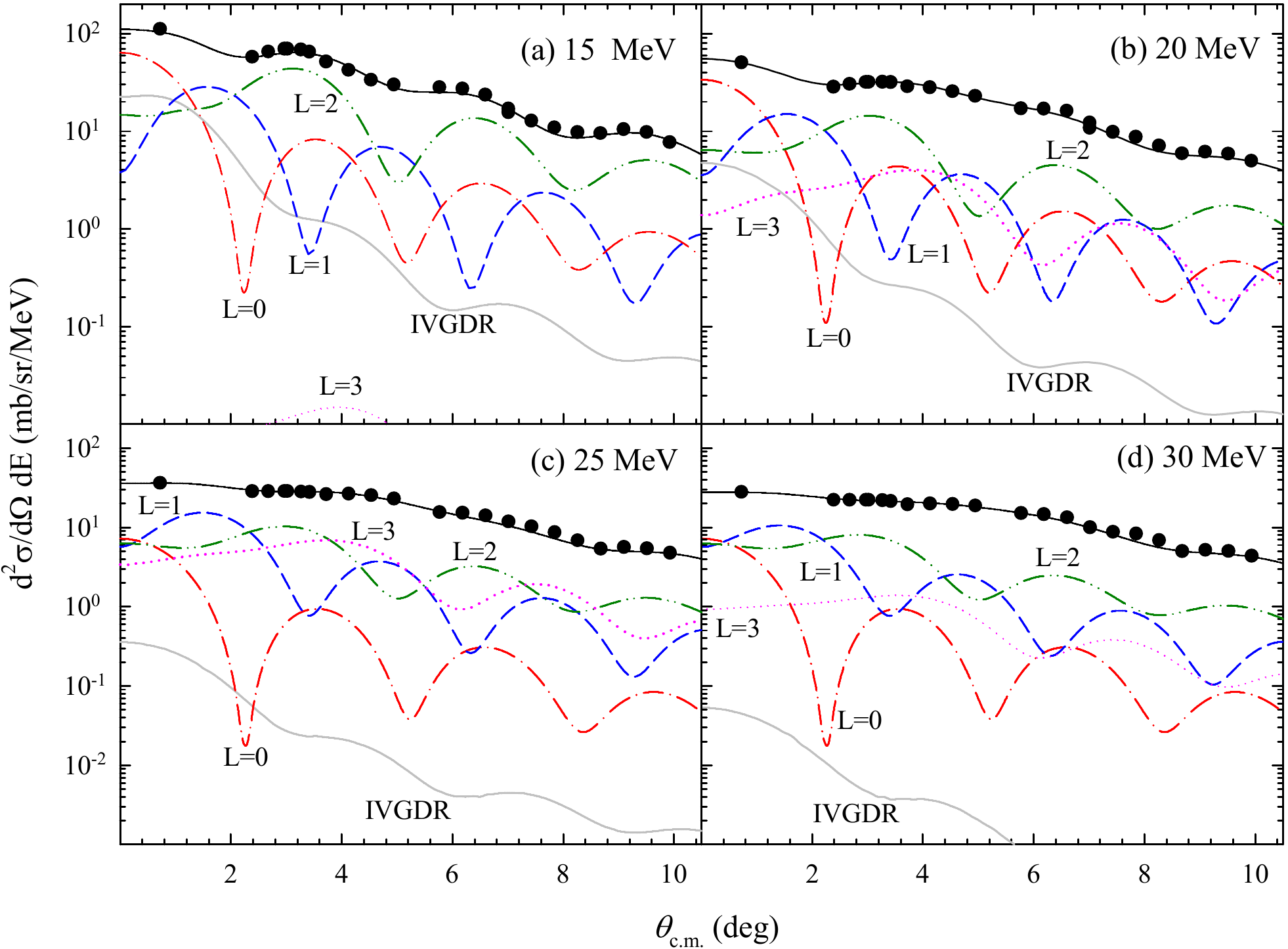}
\vspace{-0.3cm}
\caption{MDA fits to the experimental angular differential cross sections for inelastic $\alpha$-scattering data for $^{90}$Zr for several 1-MeV energy bins centered at the indicated excitation energies. The solid black solid line through the data shows the sum of various multipole components obtained from MDA. The dot-dashed (red), dashed (blue), dot-dot-dashed (green), dotted (pink), and thin solid curves (gray)
indicate contributions from $L$ = 0, 1, 2 and 3, and IVGDR, respectively. Figure from (\cite{YKG_A90_PRC}).}
\label{MDA_zr}
\end{figure*}

MDA fits for several energy bins in $^{90}$Zr are shown in Fig.~\ref{MDA_zr} for 386-MeV inelastic scattering data from RCNP (\cite{YKG_A90_PRC}); the contributions from the $L$ = 0, 1, 2, and 3 multipoles, as well as from the IVGDR are also shown.

The strength distributions for the various multipoles are obtained by multiplying the extracted $a_{L}(E_{x})$'s by the strength corresponding to 100\% EWSR at the given energy $E_x$ (\cite{Harakeh_book, Satchler1987, Harakeh1981}):
\begin{equation}
S_0(E_x)=\frac{2\hslash^2A<r^2>}{mE_x}a_0(E_x), 
\end{equation}
in the case of monopole, 
\begin{equation}
S_1(E_x)=\frac{3\hslash^2A}{32\pi mE_x}[11<r^4>-\frac{25}{3}<r^2>^2 
-10\epsilon<r^2>] a_1(E_x), 
\end{equation}
in the case of IS dipole, and
\begin{equation}
S_{L\geq2}(E_x)=\frac{\hslash^2A}{8\pi
mE_x}L(2L+1)^2<r^{2L-2}>a_2(E_x), 
\end{equation}
in the case of the higher IS multipoles.
Here, $<r^N>$ is the $N^{\rm th}$ moment of the ground-state density, while 
$\epsilon$ is defined as:
$\epsilon$~=~(4/$E_{\rm ISGQR}$+5/$E_{\rm ISGMR}$)$\hslash^2$/3$mA$. 
The centroid energies of the ISGMR and the ISGQR are generally taken as
80~A$^{-1/3}$ MeV and 64~A$^{-1/3}$ MeV, respectively.

\begin{figure}[h!]
\centering\includegraphics [height=0.50\textheight]{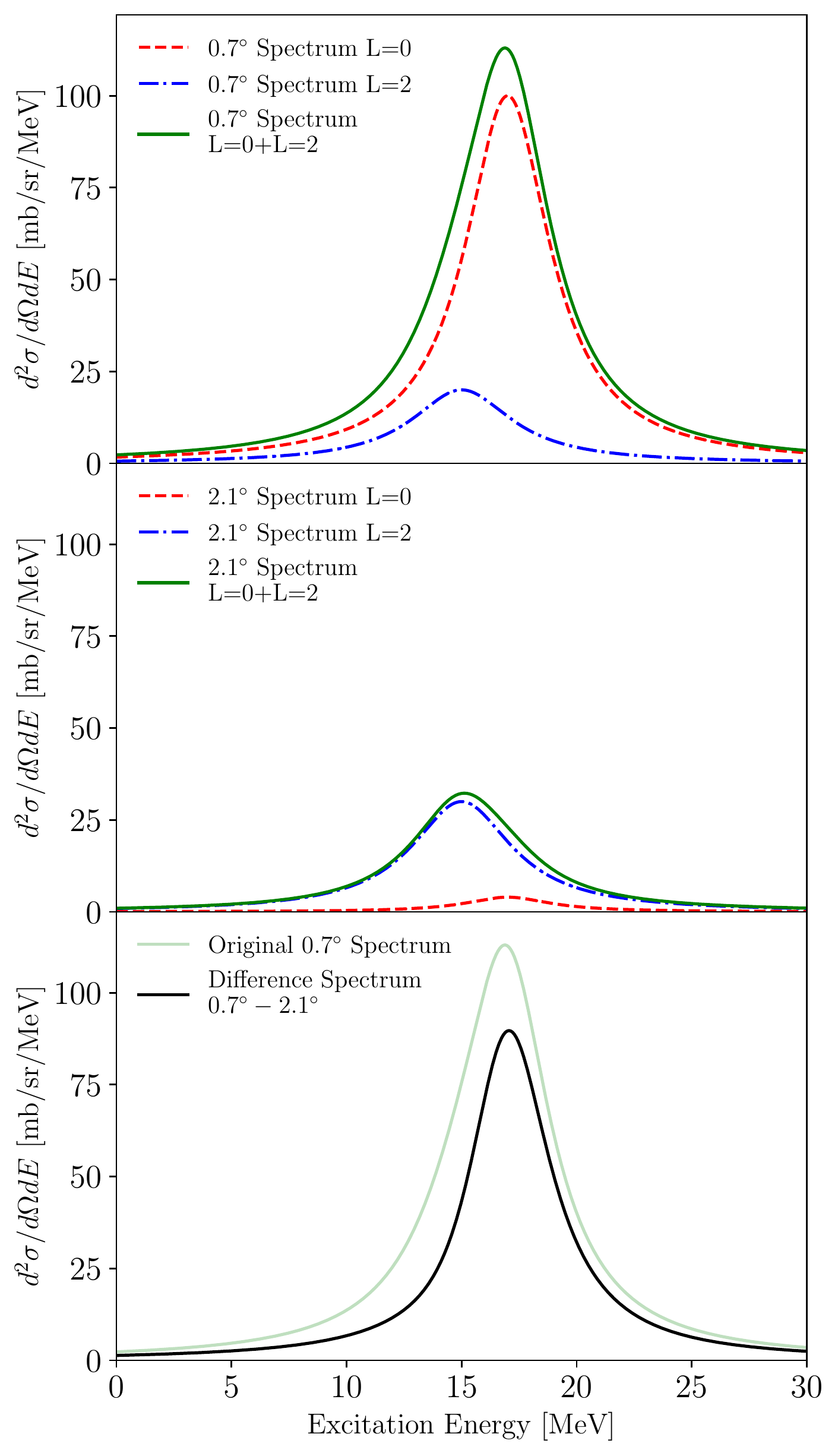}
\caption{
(a) Expected $L$=0, $L$=2 strengths for 100\% EWSR excitation in 386-MeV inelastic $\alpha$ scattering off $^{90}$Zr at 0.7$^{\circ}$. The sum of the two strengths (basically, the inelastic scattering spectrum at this angle) is also shown.
(b) Same as (a) but for 2.1$^{\circ}$. Note that the ISGMR contribution is barely discernible at this angle. (c) The ``difference spectrum'' obtained by subtracting the total ($L$=0 + $L$=2) spectrum in (b) from that in (a). The total spectrum in (c) (solid black line) is essentially the ISGMR. The total 0.7$^{\circ}$ spectrum (light green solid line) is also shown for a direct comparison.}
\label{subtr}
\end{figure}

In the absence of data at a sufficient number of angles to perform a proper multipole decomposition analysis, it is possible still to identify the position and width 
of the ISGMR and ISGDR from a ``difference of spectra'' procedure, if data are available for angles at and near 0$^{\circ}$ (see, for example, Fig. \ref{angdist} and (Brandenburg et al., 1987)). The premise behind this technique is simple: The angular distribution of 
the ISGMR is maximal at 0$^{\circ}$ and declines sharply to a minimum (see Fig. \ref{angdist}, where this minimum is at about 2$^{\circ}$). The angular distribution of the ``competing'' ISGQR, on the other hand, remains essentially flat over this angular range. Indeed, this is mostly true for all other multipoles, as well as the ``background''. So, the ``difference-spectrum'', obtained from subtracting the inelastic scattering spectrum at the first minimum of the expected ISGMR angular distribution from that at 0$^{\circ}$ (where the ISGMR strength is maximal), essentially represents only the ISGMR strength. This is shown schematically in Fig. \ref{subtr}. The same holds in comparing the

\begin{figure}[h!]
\centering\includegraphics [height=0.57\textheight]{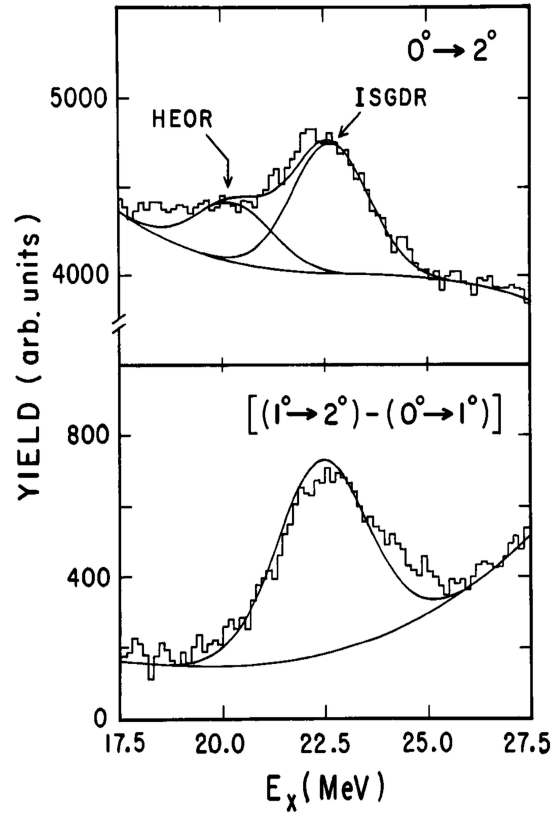}
\caption{(a) Inelastic $\alpha$-scattering spectra for $^{208}$Pb for (0$\pm$2)$^{\circ}$. A two-peak + polynomial-background fit to the data
is shown superimposed with the peaks corresponding to the HEOR and the ISGDR as indicated.
(b) The ``difference spectrum''. A fit using peak parameters identical to those in (a) is also shown; note that the fit corresponds to no HEOR strength. Figure from (\cite{benny}).}
\label{isgdr_benny}
\end{figure}

\noindent
angular distributions of the ISGDR and the ISHEOR and this technique was used, for example, in the first clear identification of the ISGDR (\cite{benny}), as shown in Fig.~\ref{isgdr_benny}. 

\section{\textit{Experimental Results}}\label{experiment_results}

\subsection{The ISGMR}
\begin{figure*}[h!]
\centering\includegraphics [height=0.35\textheight]{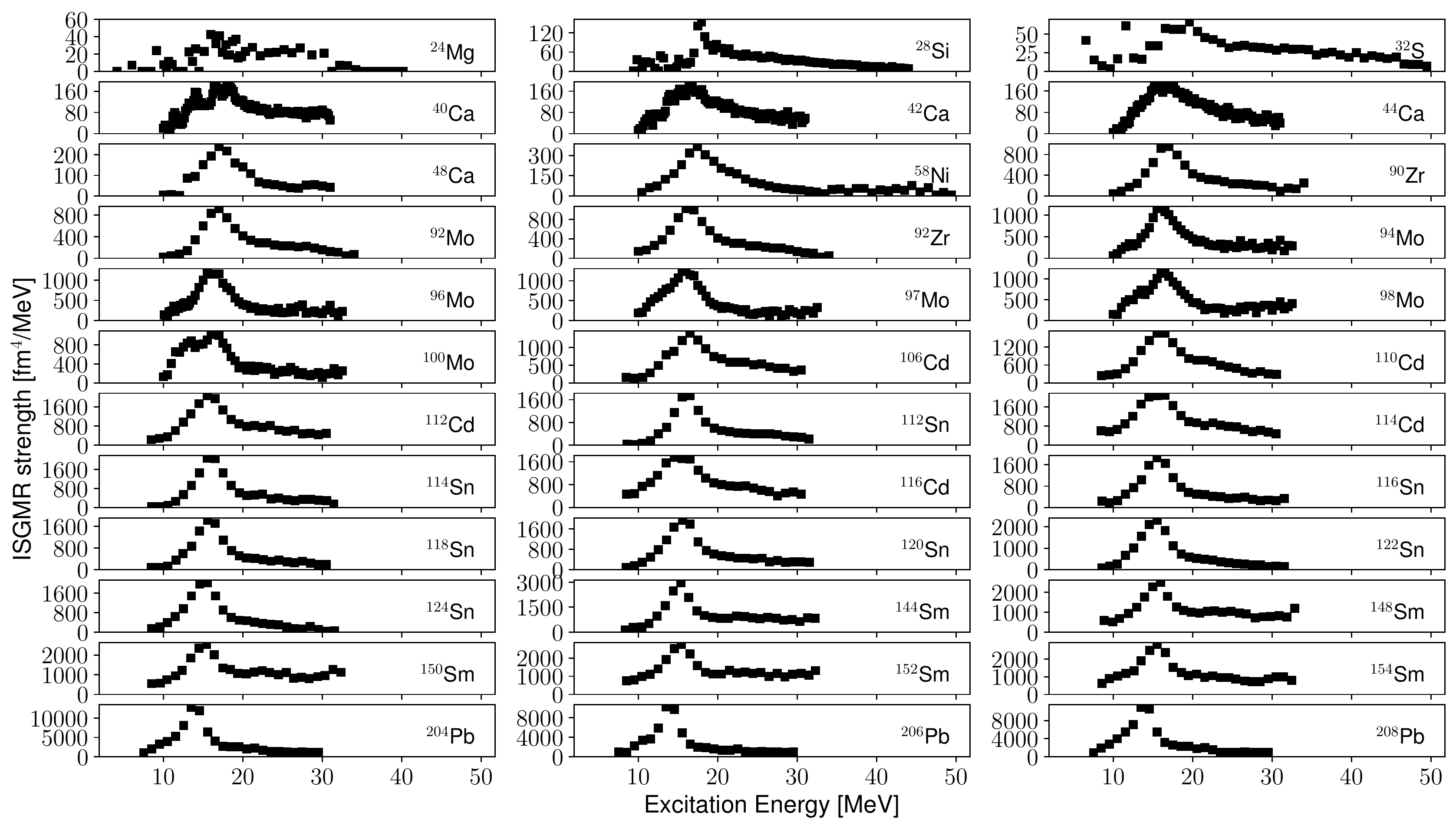}
\caption{ISGMR strength distributions for various nuclei, extracted in the RCNP work. The data are from (\cite{Itoh_prc2003,nayak58ni,Li_PRL2007,Li_2010,Darshana2012,Itoh_32S,Darshana2013,yogesh-mg2,YKGPLB2016,yogesh-mg,tom-si,KBH_CA,KBH_Mo}).}
\label{ISGMR}
\end{figure*}

Experimental strength distributions for ISGMR, based on multipole decomposition analyses, have now been obtained for several nuclei over the range A=24--208. The results from the RCNP work, from small-angle inelastic scattering of 400-MeV $\alpha$-particles are presented in Fig. \ref{ISGMR}. In practically all cases, the ISGMR appears as a single peak exhausting within it nearly 100\% EWSR for $L$=0. [In deformed nuclei, the ISGMR strength distribution has a two-peak structure (see subsection on deformation).] The associated peak parameters, as well as the various commonly-used moment ratios of the strength distributions, where available, are presented in~Table~1. The table is comprehensive in that it contains not only the results from RCNP work with $\alpha$ particles, but also their work with the deuteron probe, and the TAMU work with $\alpha$ particles and $^6$Li ions.

\begin{figure}[]
\vspace{-2.5cm}
\hspace{-4cm}
\includegraphics [width=1.7\textwidth]{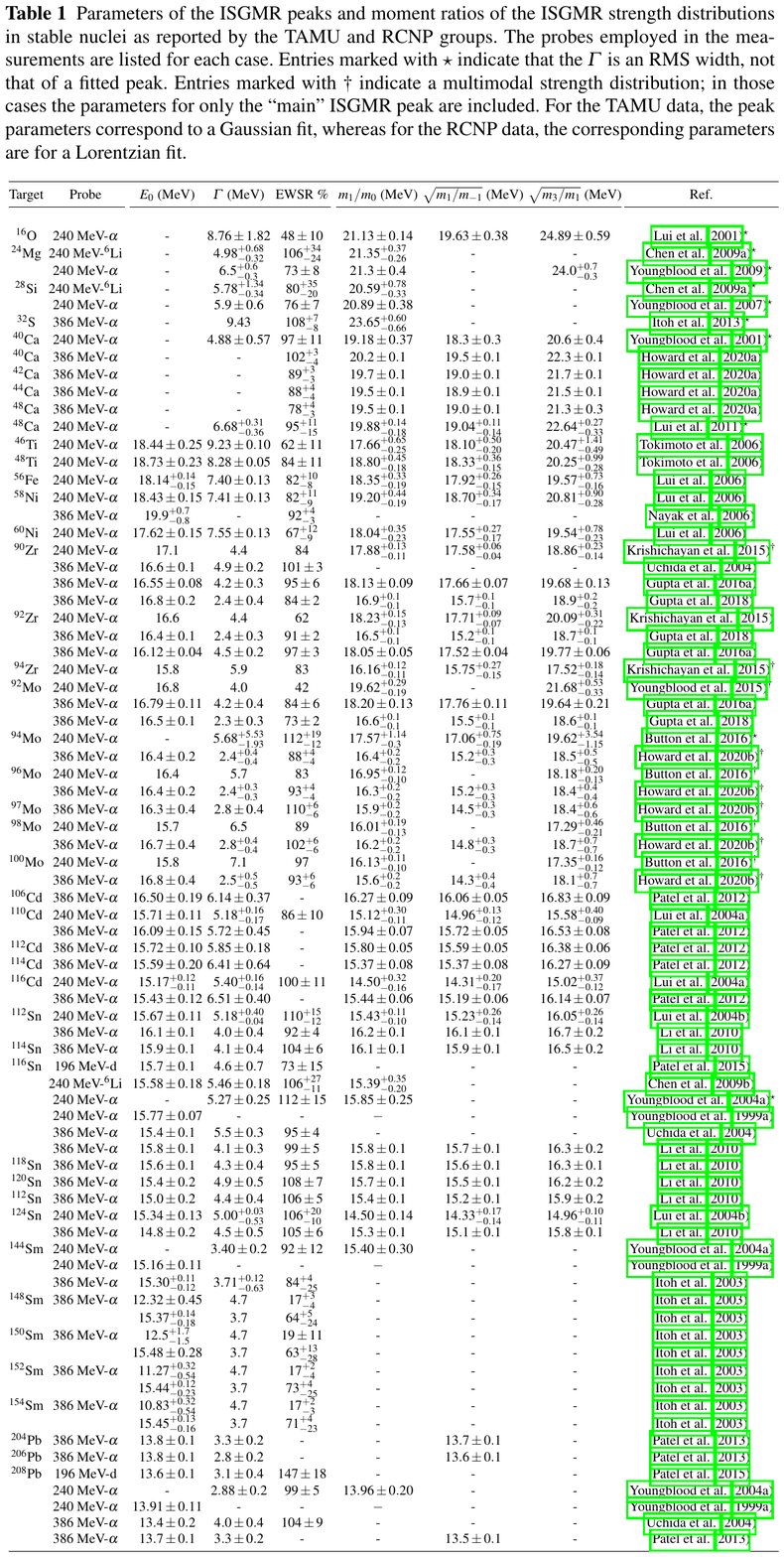}
\label{table1}
\end{figure}

A comment about the small, near-constant ISGMR strength at higher excitation energies observed in most of the RCNP work is in order. [A similar effect is observed in the ISGDR strength distributions as well (see below).] This strength is spurious and, in some ways, points to a limitation of the multipole decomposition analysis. While the correct {\em raison d'\^{e}tre} of this extra strength is not quite well understood, this may be attributed, quite reasonably, to contributions to the continuum from three-body channels, such as knockout reactions and quasi-free processes (\cite{bran1,nayak58ni}). In the MDA procedure, the continuum underlying the giant resonances is assumed to be composed of contributions from higher multipoles (hence the inclusion of multipoles up to $L$=7, typically, in the MDA). 
The aforementioned three-body processes, which also are forward-peaked in terms of angular distributions, are implicitly included in the MDA as background and may mimic the $L$=0 (and $L$=1) angular distributions, leading to such spurious multipole strengths at higher energies where the associated cross sections are very small.
This conjecture is supported by measurements of proton decay from the ISGDR 
at backward angles wherein no such spurious strength is observed in spectra in coincidence with the decay
protons (\cite{hun1,garg-rev1,hun3,nayak2}); quasi-free knockout results in protons that are forward peaked. This problem does not exist in the work of the TAMU group because, as mentioned previously, they subtract all ``background'' from their spectra before carrying out MDA. However, a similar
increase in the ISGMR strength at high excitation energies was reported as well in their results 
in $^{12}$C when they carried out MDA {\em without} first subtracting the continuum from the excitation-energy spectra (\cite{john}).

\subsection{The ISGDR}
\begin{figure*}[!h]
\centering\includegraphics [height=0.35\textheight]{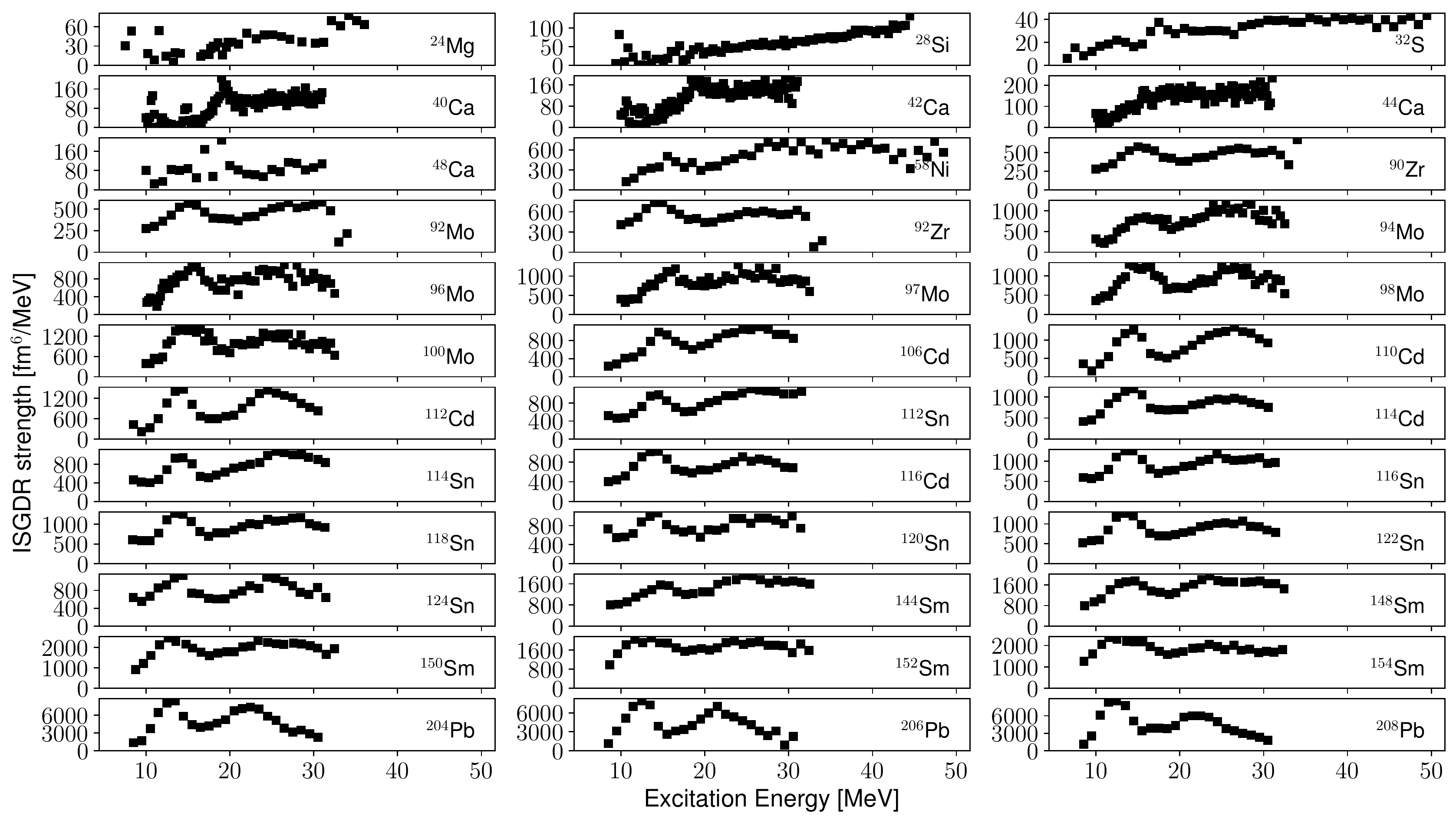}
\caption{ISGDR strength distributions for various nuclei, extracted in the RCNP work. The data are from (\cite{Itoh_prc2003,nayak58ni,Li_PRL2007,Li_2010,Darshana2012,Itoh_32S,Darshana2013,yogesh-mg2,YKGPLB2016,yogesh-mg,tom-si,KBH_CA,KBH_Mo}).}
\label{ISGDR}
\end{figure*}

Experimental ISGDR strength distributions extracted in the RCNP work are presented in Fig. \ref{ISGDR}. 
The properties of the ISGDR peaks, in nuclei where a distinct peak structure is observed, are summarized in Table~2. This table is also comprehensive in that it includes all available results from both RCNP and TAMU.

\begin{figure}[]
\hspace{-1cm}
\includegraphics [height=1.02\textheight]{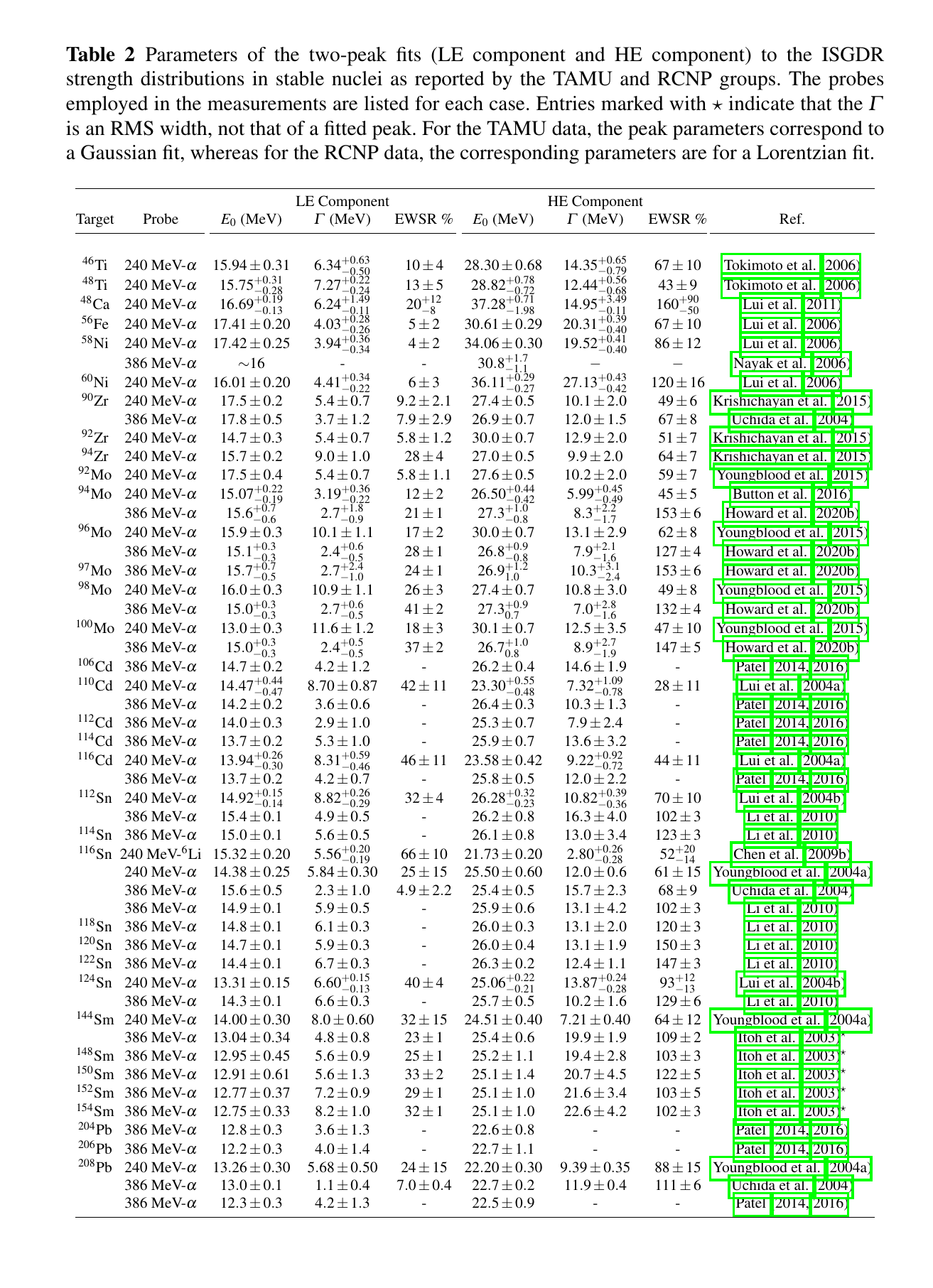}
\label{table2}
\end{figure}

Two points need to be made in this connection: i) No ``peak'' structure is observed for ISGDR in the light nuclei, and generally one sees a monotonously increasing ISGDR strength at excitation energies above ~20 MeV. This is believed to be because of the aforementioned ``spurious strength distributions'' observed also for the ISGMR; this effect is further exacerbated by the fact that the peak energy of the ISGDR itself is rather high ($>$25 MeV) in the light nuclei.
ii) In all heavy nuclei (A$\geq$58), the ISGDR strength distribution has two distinct peaks. 
This ``low-energy'' (LE) isoscalar $L$=1 strength has engendered considerable
interest and argument. It is present in nearly all of the recent theoretical calculations in some form or the other, and at similar energies, although with varying strength. It has been shown(\cite{co00,jorge2000,dv00,shlomo2002}) that the centroid of this component
of the $L$=1 strength is independent of the nuclear incompressibility (and, hence, is certainly of ``non-bulk'' nature). While the exact nature of this component is not fully
understood yet, suggestions have been made that this component might
represent the ``toroidal''~\cite{dv00,balb} or the ``vortex'' modes~\cite{dubna1,dubna2}. 
It is impossible to distinguish between the competing possibilities based on currently-available
data~\cite{garg-rev1}; also, it is not at all clear why these exotic modes would be excited with such large cross sections in ($\alpha,\alpha'$) work. There is general agreement, however, that only the
high-energy (HE) component of this bi-modal distribution needs to be
considered in obtaining a value of $K_A$ from the energy of the ISGDR.S
Nearly all the expected $L$=1 EWSR strength is observed under this HE peak in all cases.

Incidentally, the ISGMR and ISGDR data in $^{208}$Pb give a consistent value for 
the finite-nucleus incompressibility $K_A$ and, hence,
$K_\infty$ (\cite{Uchida_PLB2003,garg-rev1}). 

\subsection{The ISGQR}
\begin{figure*}[h!]
\vspace{-0.5in}
\centering\includegraphics [height=0.45\textheight]{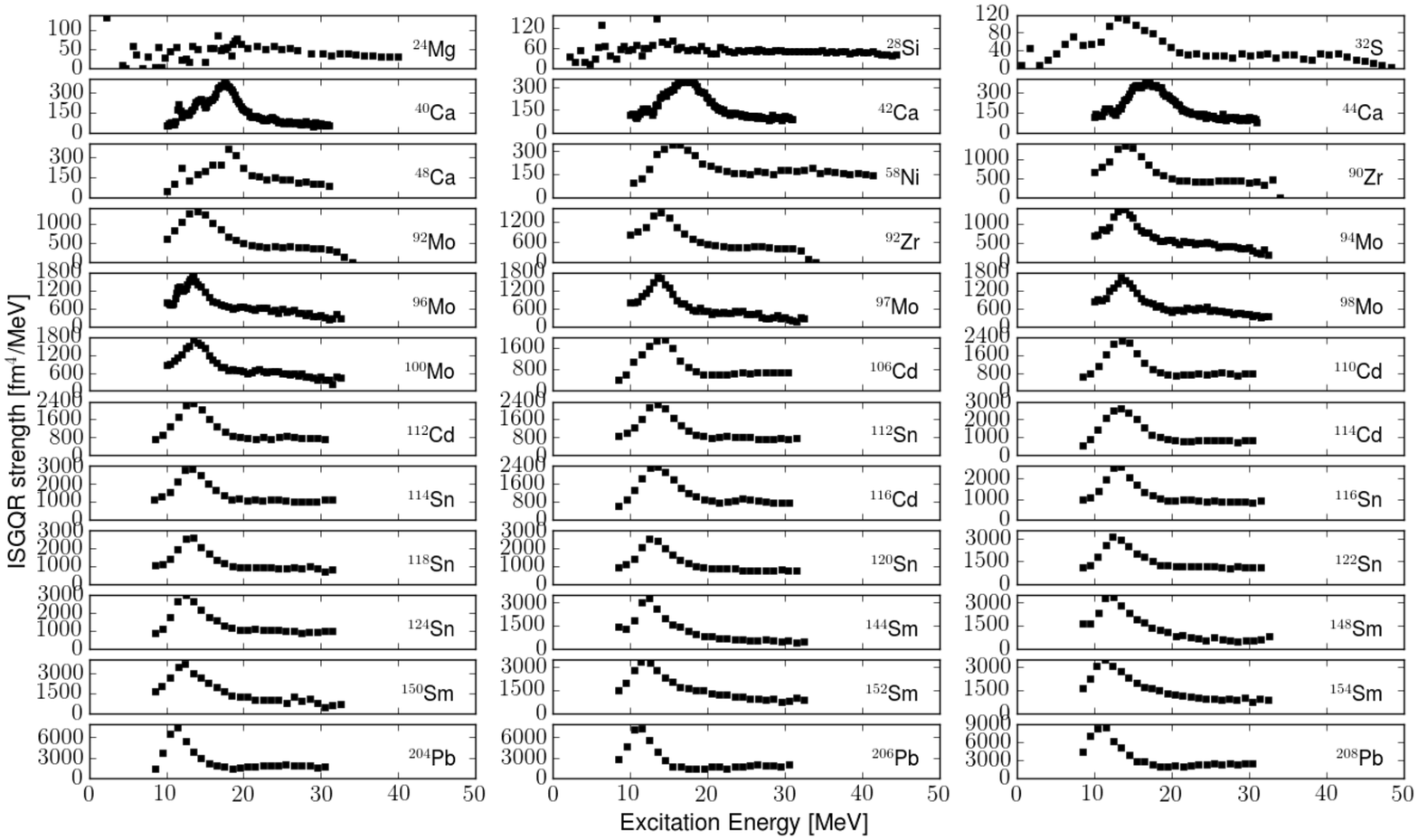}
\vspace{-0.5in}
\caption{ISGQR strength distributions for various nuclei, extracted in the RCNP work. The data are from (\cite{Itoh_prc2003,nayak58ni,Li_PRL2007,Li_2010,Darshana2012,Itoh_32S,Darshana2013,yogesh-mg2,YKGPLB2016,yogesh-mg,tom-si}).}
\label{ISGQR}
\end{figure*}

Experimental ISGQR strength distributions extracted in the RCNP work are presented in Fig. \ref{ISGQR}. 
The properties of the ISGQR peaks, in nuclei where a distinct peak structure is observed, are summarized in Table~3. The results from TAMU measurements are generally consistent with the RCNP results.

\subsection{The ISHEOR}
\begin{figure*}[!h]
\centering\includegraphics [height=0.38\textheight]{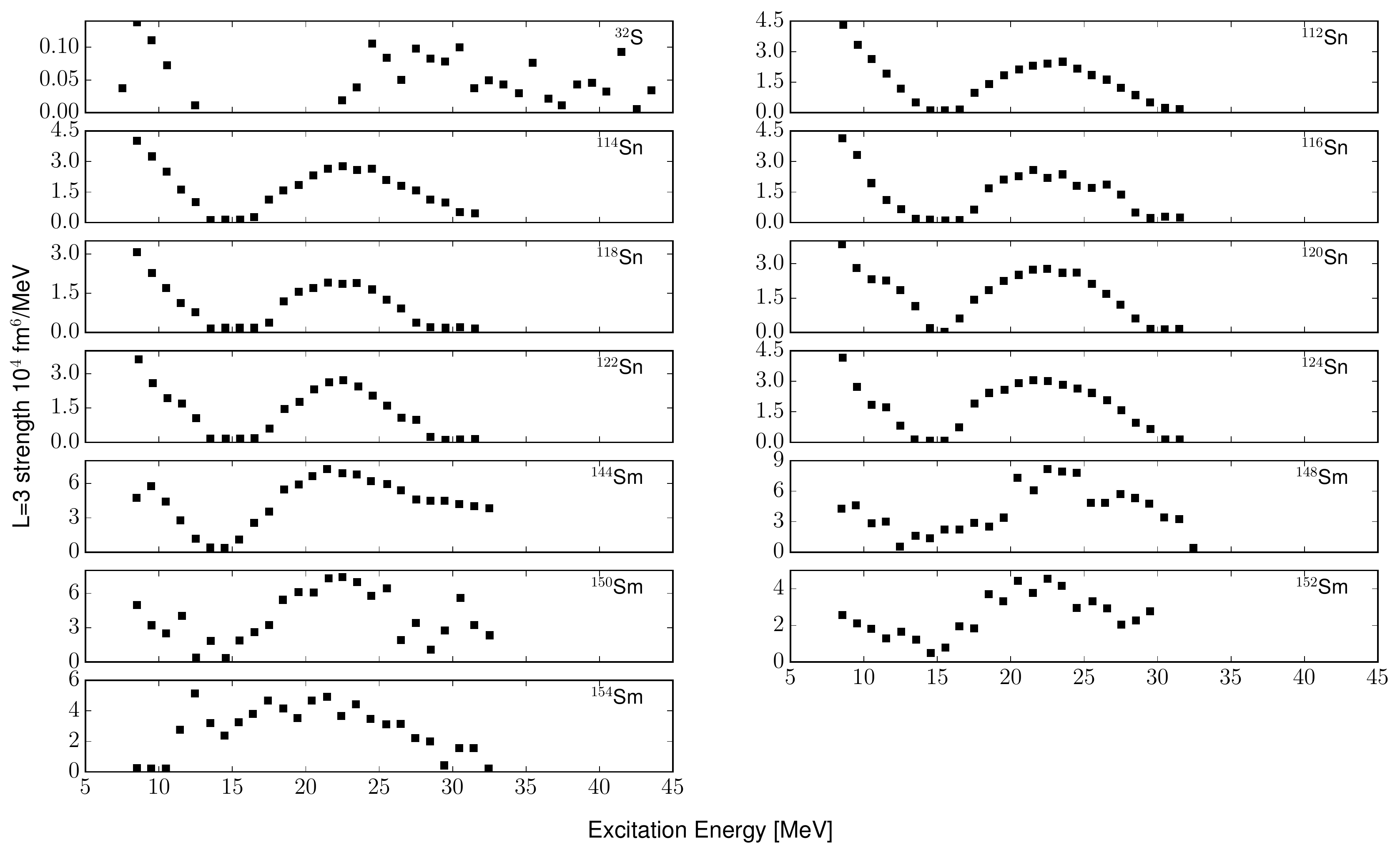}
\caption{ISHEOR strength distributions for various nuclei, extracted in the RCNP work. The data are from (\cite{Itoh_prc2003,Li_PRL2007,Li_2010,Itoh_32S}).}
\label{HEOR}
\end{figure*}

Experimental ISHEOR strength distributions extracted in the RCNP work are presented in Fig. \ref{HEOR}. 
The properties of the ISHEOR peaks, in nuclei where a distinct

\noindent
peak-structure is observed, are summarized in Table~4. The results from TAMU measurements, where available, are generally consistent with the RCNP results and are also included in Table~4.
As stated before, the $L$=3 resonances have another component, the 1$\hslash\omega$ ISLEOR. This resonance, previously identified in several measurements [for example, (\cite{moss1976} and \cite{morsch1980})] is at an excitation energy too low to be fully observed in the more recent work, although the high-energy side of that component is apparent in almost all cases at the low-excitation-energy part of the strength distributions presented in Fig. \ref{HEOR} and, in some cases, it has been possible to extract tentative parameters for the ISLEOR from fits to the low-excitation-energy parts (see Table~4).

\begin{figure}[!ht]
\centering
\includegraphics [height=0.90\textheight]{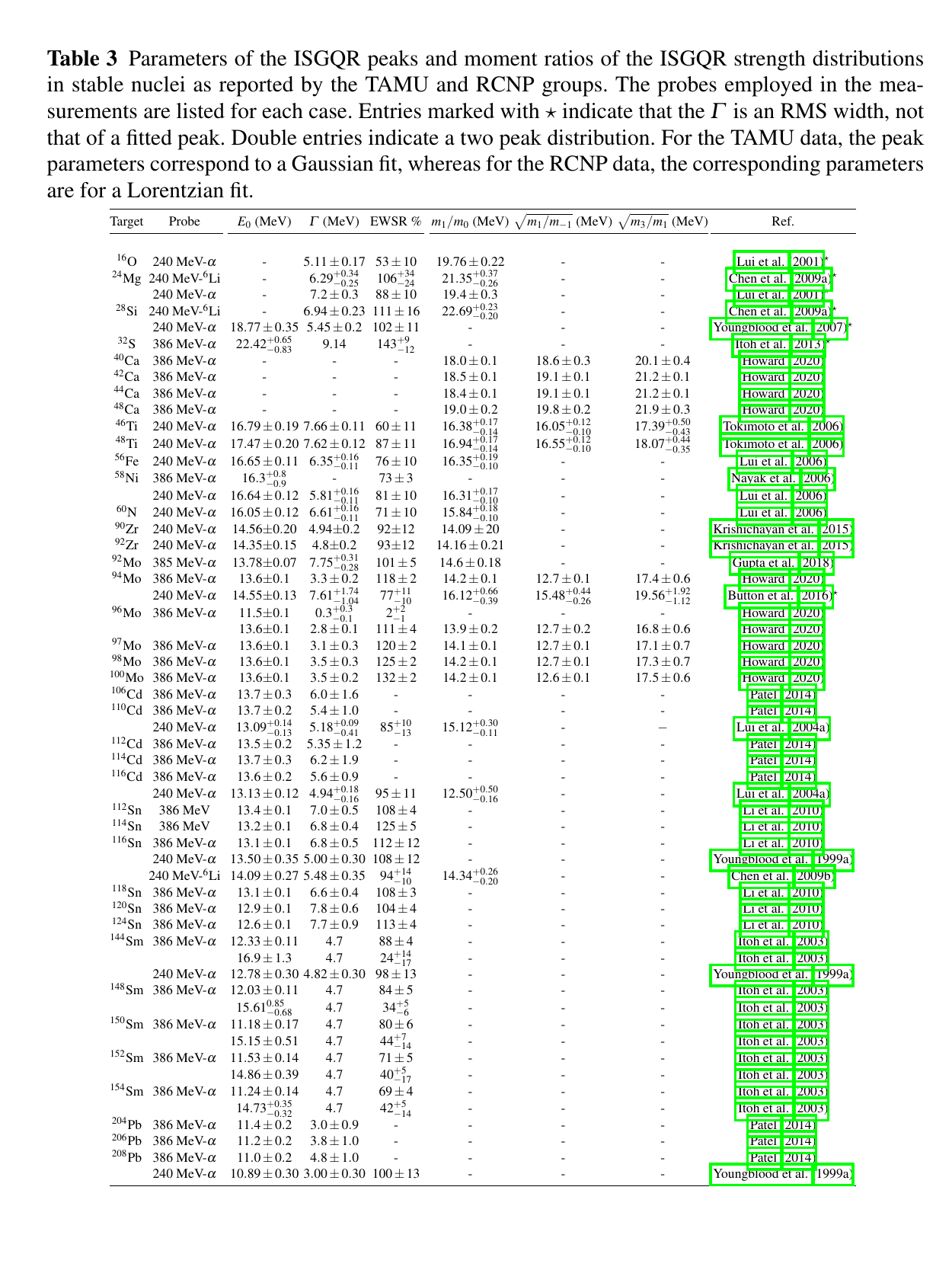}
\label{table3}
\vspace*{-2cm}
\end{figure}

\begin{figure}[!ht]
\centering
\includegraphics [height=0.8\textheight]{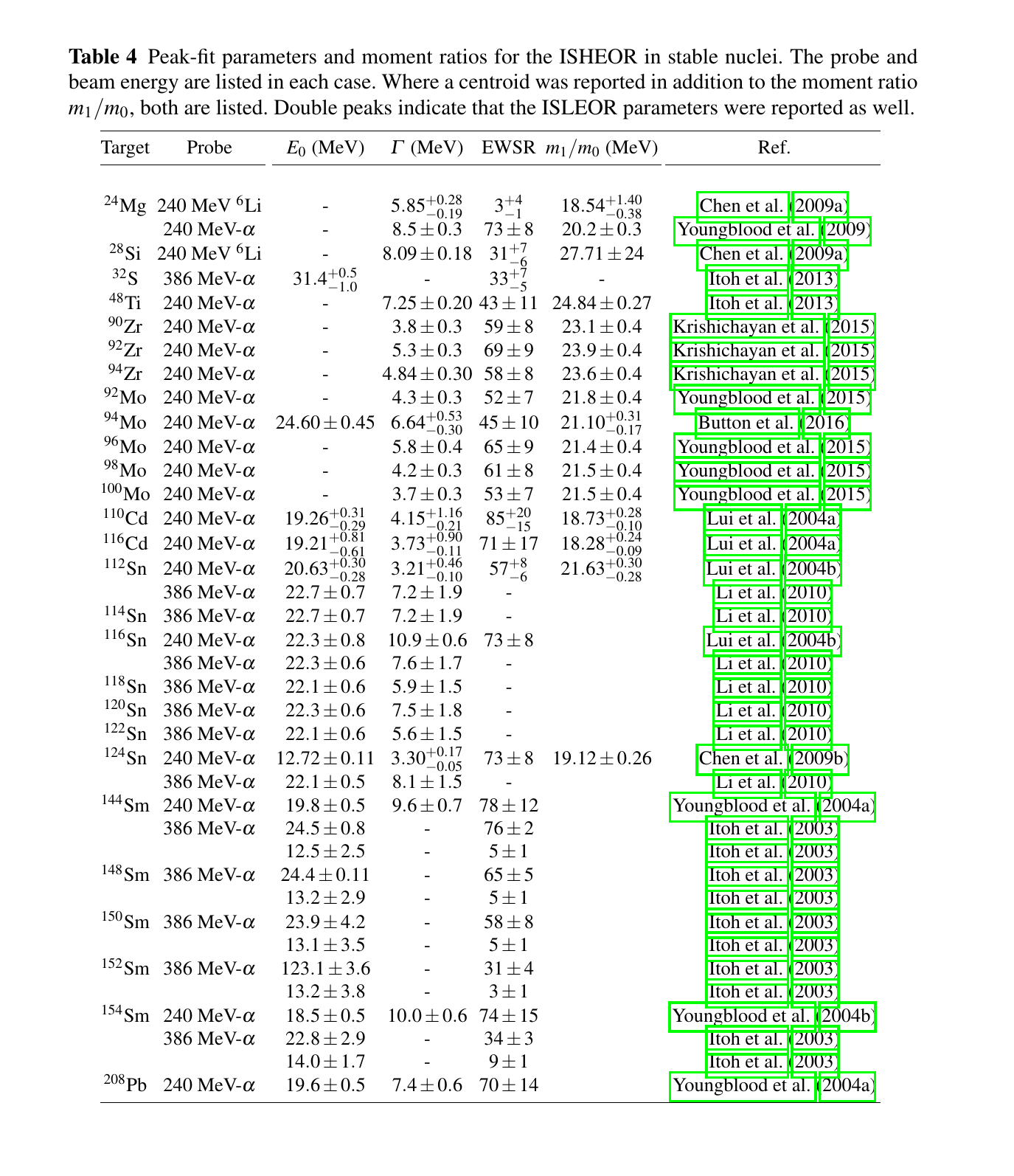}
\label{table4}
\vspace*{-1cm}
\end{figure}

\section{\textit{Deformation Effects on Isoscalar Giant Resonances}}

So far we have discussed the isoscalar giant resonances only in the ``spherical'' nuclei. The structure of the giant resonances is affected by the deformation of the nuclear ground state. This effect for the IVGDR has been known for a long time: that resonance splits into two components corresponding to the different frequencies of dipole oscillations along the major and minor axes of a symmetric ellipsoidal nuclear shape (\cite{Harakeh_book}). The ISGQR, on the other hand, exhibited only a small broadening due to deformation of the ground state: the $L$ = 2 resonance splits into three components corresponding to the $K$ quantum numbers $K$ = 0, 1, and 2. These components are rather closely-spaced leading to an overall increase in ISGQR width in a deformed nucleus ($^{154}$Sm, for example), as compared with that in a spherical nucleus ($^{144}$Sm) (\cite{kishimoto}); calculations with a modified and self-consistent quadrupole-quadrupole interaction produced a splitting of the ISGQR in $^{154}$Sm consistent with the experimental data.

\begin{figure}[!h]
\vspace*{-0.2cm}
\centering\includegraphics [height=0.50\textheight]{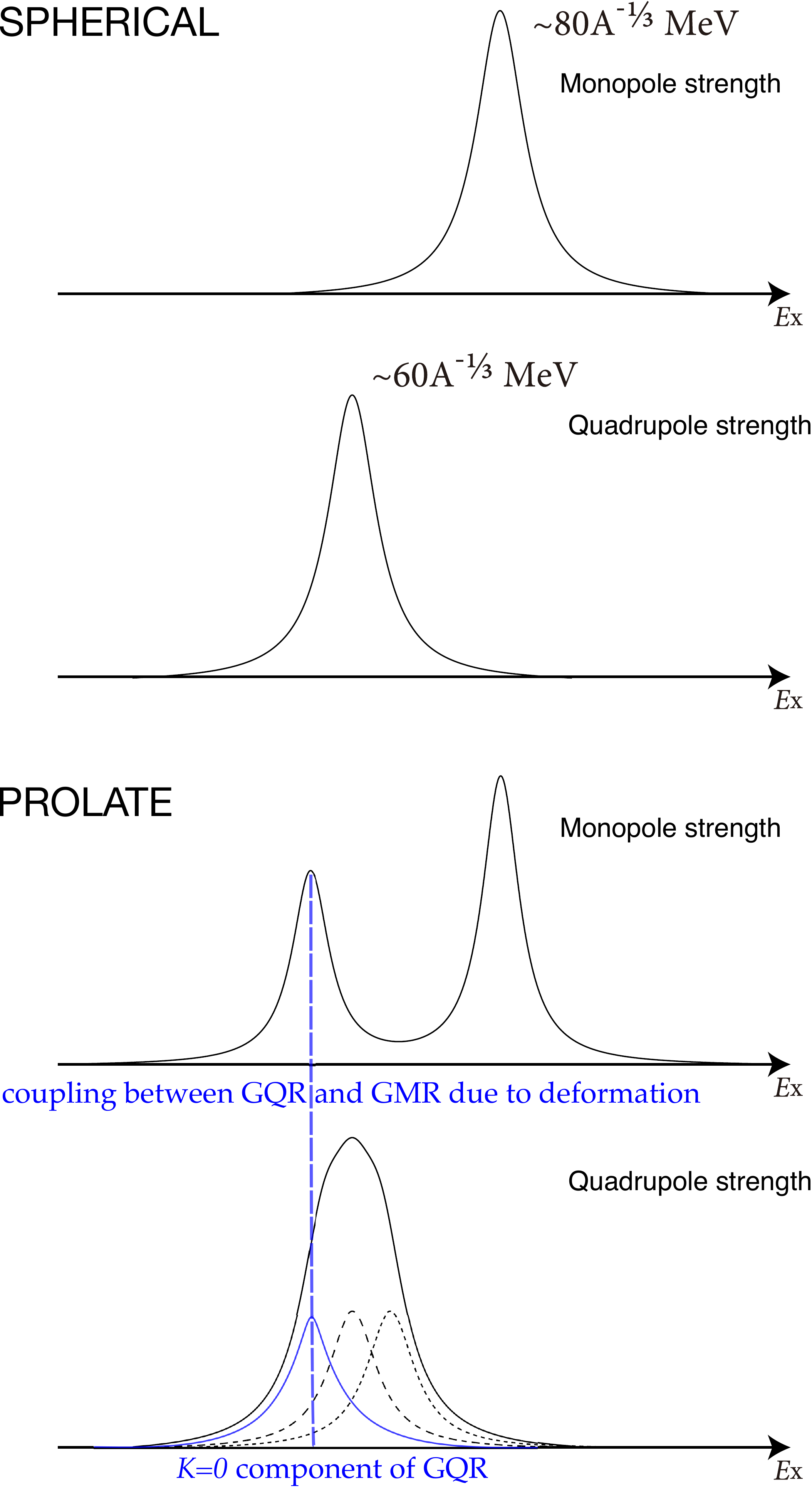}
\caption{Schematic representation of the effect of ground-state deformation on the ISGMR and ISGQR. Figure courtesy of Dr. K. Yoshida, Kyoto University, Japan.}
\label{splitting}
\end{figure}

Naively, one would have expected the $L$ = 0 ISGMR to remain unaffected by the deformation of the ground state; however, one does see a ``splitting'' of the ISGMR strength in deformed nuclei, as first reported by Garg et al. in $^{154}$Sm (\cite{umesh_prl2}). This ``splitting'' results from a coupling of the $K$ = 0 component of the ISGQR with the ISGMR. The monopole and quadrupole vibrations in the deformed nuclei no longer have a unique $J^\pi$, each containing a mixture of $L$ = 2 and $L$ = 0 instead. Thus, there are two $K$ = 0 states, the lower predominantly $L$ = 2, but containing significant $L$ = 0 strength; the upper predominantly $L$=0 but with a small amount of $L$ = 2 strength (\cite{umesh_prl2}). This is represented schematically in Fig.~\ref{splitting}. A similar splitting was predicted in calculations in the framework of the 

\noindent
quasi-particle random-phase approximation (QRPA) (\cite{Speth1978}) and in the adiabatic cranking approximation (\cite{abgrall}).

\begin{figure}[h]
\centering\includegraphics [height=0.60\textheight]{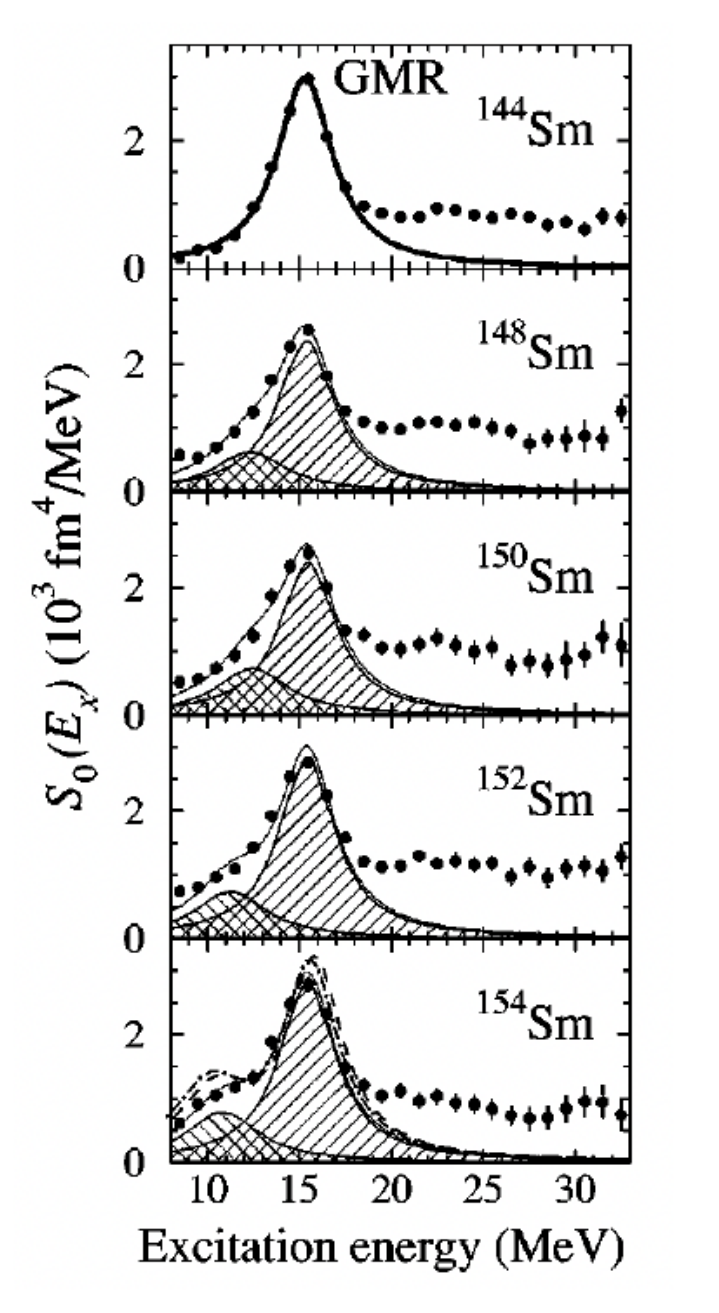}
\caption{Inelastic $\alpha$-scattering spectra at $\theta_{\rm av}$ = 0.7$^{\circ}$ and $E_\alpha$ = 386 MeV for $^{144-154}$Sm. Figure from (\cite{Itoh_prc2003}).}
\label{isgmr_sm}
\end{figure}

In the Sm isotopes, which range from the ``spherical" $^{144}$Sm (deformation parameter $\beta_2$ = 0.09) to the well-deformed $^{154}$Sm ($\beta_2$ = 0.34), the evolution of the ISGMR strength as a function of increasing deformation is observed rather succinctly (\cite{Itoh_prc2003}): a single peak in $^{144}$Sm evolves into two clearly discernible components in case of $^{154}$Sm  (see Fig.~\ref{isgmr_sm}). Indeed, this evolution is evident even in the ``0$^{\circ}$'' inelastic scattering spectra [see Fig. 4 of (\cite{Itoh_prc2003})]. A clear two-component structure in the ISGMR strength distribution was reported in the work of the TAMU group as well (\cite{dhybg,dhy-sm2}). A splitting of the monopole strength in the deformed nuclei was also observed in the ($^3$He,$^3$He$'$) work (\cite{Buenerd1980}).

The effect of deformation on ISGDR is similar in that there is coupling between the $K$ = 1 (as well as the $K$ = 0) components of the ISGDR ($L$ = 1) and the ISHEOR ($L$ = 3) (\cite{Ando1985}). However, because of the aforementioned LE component of the ISGDR even in the spherical nuclei, this coupling is not as clearly evident as in case of the ISGMR. However, two effects are discerned in going from $^{144}$Sm to $^{154}$Sm, both consistent with the coupling between the $K$ = 1 components of the two resonances (\cite{Itoh_plb,Itoh_prc2003}): i) the relative strength of the LE component of the ISGDR increases smoothly with nuclear deformation, whereas the strength of the HE component remains constant; ii) the width of the LE component also increases with increased deformation. A direct comparison of the ISGDR strength in the deformed nucleus, $^{154}$Sm, is complicated, of course, by the uncertain nature of the LE component of the isoscalar $L $ = 1 strength distribution.

A most interesting result on the effect of deformation on the ISGMR strength was observed in the nucleus $^{24}$Mg (\cite{yogesh-mg2,yogesh-mg}). Generally, the ISGMR strength (indeed, all multipole strengths) in the lighter-mass nuclei (A$<$58) is fragmented over a wide excitation-energy range and does not form a nice ``peak'' as in the heavier nuclei (see, for example, (\cite{lui2001,Itoh_32S})). With that, any effects of deformation 
were expected to be very difficult to discern. However, in RCNP work on this deformed nucleus, a two-peak structure was observed in the ISGMR strength distribution, indicative of the ``splitting'' of the ISGMR. The observed strength distribution is in good agreement with microscopic calculations for a prolate-deformed ground state for $^{24}$Mg, carried out in a deformed Hartree-Fock-Bogoliubov (HFB) approach and QRPA with a Skyrme and Gogny energy-density functional (\cite{Peru2008,Yoshida2010}), and is in contrast with what

\begin{figure}[!h]
\centering\includegraphics [height=0.35\textheight]{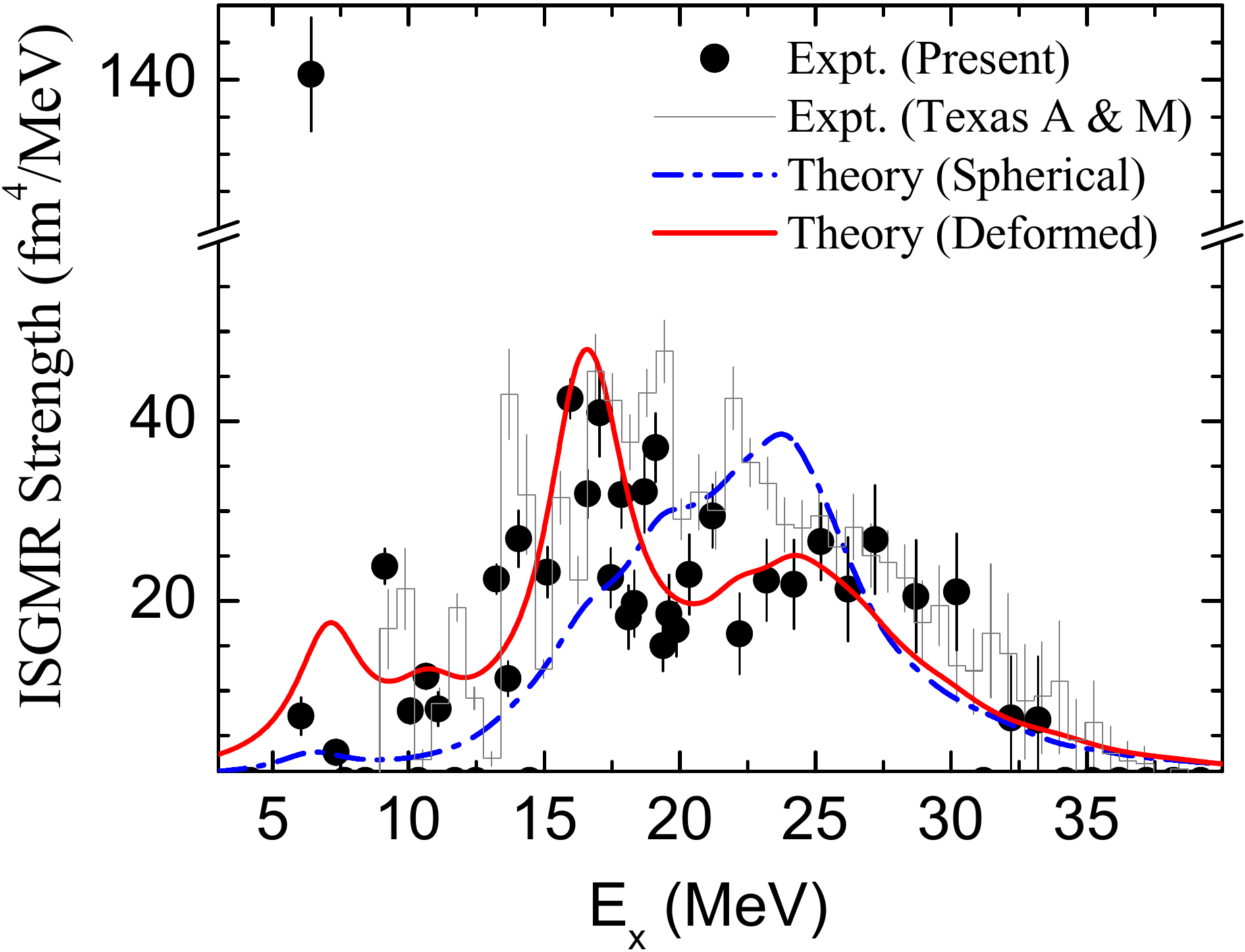}
\caption{ISGMR strength distribution in $^{24}$Mg (solid circles). The dash-dotted (blue)  and solid (red) lines  show microscopic calculations for spherical and prolate ground-state deformation, respectively. Figure from (\cite{YKGPLB2016}).}
\label{mg24}==
\end{figure}
\noindent
would be expected if a spherical ground state is assumed for this nucleus (see Fig.~\ref{mg24}). Another set of calculations, with the SkM*, SVbas, and SkP$^{\delta}$ interactions, further confirms that the $E$0 peak at $E_x\sim$16 MeV is caused by the deformation-induced coupling of the ISGMR with the $K$ = 0 part of the ISGQR (\cite{kvasil}).
\noindent
A similar effect has since been observed in the oblate-deformed nucleus $^{28}$Si as well (\cite{tom-si}).
The effect of deformation on ISGMR and ISGQR has also been investigated theoretically in the molybdenum isotopes, using a fully-consistent QRPA approach with 
several Skyrme interactions (\cite{Colo2020}).

\section{\textit{Measurements of Decays of Giant Resonances}}

Very important information on the microscopic nature of giant resonances, in particular on the particle-hole states involved, can come from measurements of the charged-particle and neutron decays of the resonances. Since these resonances are located well above the particle-separation thresholds, the dominant process that  takes place is particle emission. It can occur either from the initial 1p-1h state, leaving a single-hole state in the A$-$1 nucleus (the direct component), or from the states with partially or completely equilibrated configurations, resulting in an evaporation-like spectrum of the emitted particles (the statistical component). The observation of particle emission spectra observed in coincidence with the excitation of giant resonances can provide useful information on the evolution of the decaying configuration and on the microscopic structure of the resonances (\cite{Harakeh_book,hun4}).
Furthermore, coincidence with particle decay gives a useful means to suppress the backgrounds in inelastic-scattering spectra, which can help isolate the resonance strengths more reliably. Indeed, such measurements have been very effective in eliminating both the instrumental background and the non-resonant continuum, overcoming, in the process, the problems connected with background subtraction discussed earlier, and allowing for a determination of the giant-resonance gross properties with better precision (\cite{Harakeh_book,hun4}).

Several decay measurements were carried out in the 1980's. These focused on the  particle decay of ISGMR and ISGQR. While charged-particle decay occurs preferentially in the light nuclei (A$\lessapprox$60), neutron emission is the dominant decay mode for the medium-to-heavy mass nuclei because of Coulomb barrier effects. As stated earlier, giant resonances have been observed as a single peak only int the heavier nuclei; so, the available decay mode in these measurements is primarily neutron decay [there also is fission decay in the actinide nuclei (\cite{fission})]. In most cases, the observed decay had a large statistical component, making the separation of the direct decay component an experimentally challenging task: both the resonance and the underlying continuum decay in the same manner. Still, an excess of neutron emission over the statistical-model calculations was identified for the ISGMR in some cases 
[see, for example, the measurements reported in Refs. (\cite{bran1,Bracco1989}), and the corresponding theoretical calculations (\cite{Colo1992})].

The situation turns out to be somewhat different (in the positive sense) for the ISGDR, where significant decay widths from the resonance in $^{208}$Pb to specific proton-hole states in $^{207}$Tl was predicted by continuum random-phase approximation (CRPA) calculations (\cite{urin1,urin2}); a
similar behavior was predicted in other medium- and heavy-mass nuclei. Charged-particle decay measurements are, in general, easier than their neutron-decay counterparts and the predicted decay branching ratios were sufficiently large to make such measurements quite feasible within the beam times typically granted for an experiment at international accelerator facilities.

There have been several investigations of the proton-decay of the ISGDR in recent years. At KVI, Groningen, measurements were performed on $^{58}$Ni, $^{90}$Zr, $^{116}$Sn,

\begin{figure}[!h]
\centering\includegraphics [height=0.55\textheight]{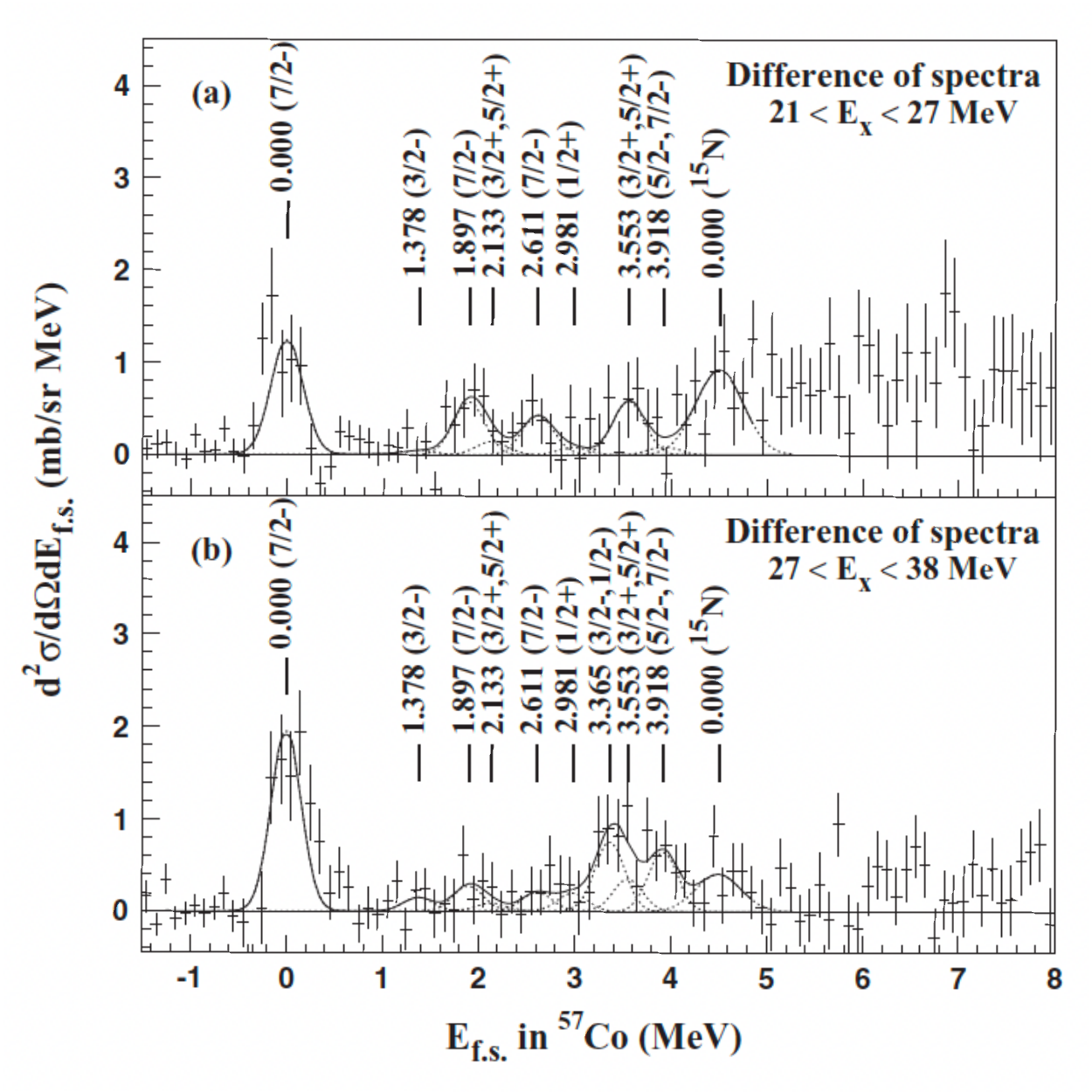}
\caption{Double-differential cross sections from difference final-state spectra (generated by subtracting the final-state spectra taken at $\theta_{c.m.}$ = 1.8$^{\circ}$--2.5$^{\circ}$ and $\theta_{c.m.}$ = 3.1$^{\circ}$--3.8$^{\circ}$) plotted as function of the excitation energy in $^{57}$Co. Coincidence conditions are set for (a) low and (b) high-energy components of the ISGDR strength in $^{58}$Ni. Figure from (\cite{hun4}).}
\label{ni_decay}
\end{figure}
\noindent
and $^{208}$Pb, using 200-MeV $\alpha$-particle beams provided by the AGOR superconducting cyclotron facility (\cite{hun1,hun3,hun4}). At RCNP, similar measurements were carried out on $^{208}$Pb, using 400-MeV $\alpha$ particles (\cite{nayak2}). In all cases, significant decay was observed to specific particle-hole states in the daughter nuclei, in qualitative agreement with the predictions of the theory; indeed, the agreement with theory was termed excellent for $^{208}$Pb (\cite{hun1,hun3,nayak2}). An example of the quality of the decay spectra from the KVI measurement on $^{58}$Ni is provided in Fig.~\ref{ni_decay}.

\section{\textit{Isoscalar Giant Resonances with Rare-Isotope Beams}}

As more and more radioactive ion beams become available, the investigation of the isoscalar giant resonances in nuclei far from stability becomes extremely interesting in order to understand and delineate the effect of large neutron-proton asymmetries on the structure of the resonances. In particular, investigations of the compression-mode resonances -- ISGMR and ISGDR -- are especially pertinent because of their direct connection to the nuclear incompressibility. For one, the effect of the asymmetry term in going from the incompressibility of individual nuclei, $K_A$, to incompressibility of infinite nuclear matter, $K_\infty$ is quite important but not well understood. There also is the intriguing possibility of the observation of the soft GMR, akin to the soft giant dipole resonance (the so-called pygmy dipole resonance) observed in the halo nuclei (see the chapters by Lanza and Vitturi, and Zilges and Savran). Thus, one would be looking at two nuclear incompressibilities: one for the core, and the other for the ``halo'' or the ``skin''. 

Calculations for nuclei far from the stability line (\cite{Hamamoto:1997,Sagawa1998}) have indicated a threshold effect in the monopole response resulting in considerable ISGMR strength at low energies. Similar effects have been  predicted in calculations by (\cite{Khan2011}) and \cite{Piekarewicz2017}) as well.  Although the conclusions of (\cite{Khan2011}) were later questioned in (\cite{Sagawa2014,jorge2015}), the nature of the ISGMR strength in nuclei far from the stability line remains of tremendous current interest. 

These measurements have to be performed in the inverse-kinematics mode with the concomitant problem of very low velocities of the recoiling target nuclei at forward-angles essential for identifying the ISGMR with multipole-decomposition analysis. As stated before, the best experimental probes for the investigation of the ISGMR are deuterons and $\alpha$ particles. For the excitation-energy range corresponding to the ISGMR, the expected energies of the recoiling particles is in the range of $\sim$100 keV to $\sim$2 MeV. In a standard thin target and particle telescope set-up, this energy would necessitate use of very thin targets ($\sim$100 $\mu$g/cm$^2$) and detectors that have practically no dead-layers or entrance foils. Considering the intensities available for radioactive ion beams, these requirements make such measurements practically impossible at this time.

The one reasonable option at present is the use of an active-target timing projection chamber (AT-TPC). In an active-target system, the detector gas employed in the TPC also acts as the target. Such a system, in principle, can have an angular coverage close to 4$\pi$, a low-energy threshold, and large effective target thickness, alleviating all the problems mentioned above associated with the inverse kinematics measurements with radioactive ion beams (\cite{mittig-ahn}). 

The first such experiment, meant primarily to establish the technique, was performed at the GANIL facility in France, using the AT-TPC system MAYA (\cite{maya}). A $^{56}$Ni beam at energy of 50 MeV/nucleon was incident on MAYA filled with deuterium gas at a pressure of 1050 mbar which is equivalent to a pure deuterium target of 1.6 mg/cm$^2$-thickness (\cite{ni56}). Even with an effective data-taking time of only 15 hours and an average beam intensity of 5$\times$10$^4$ pps, it was possible to observe the ``bump'' corresponding to the ISGMR+ISGQR in the spectrum of recoil deuterons. Moreover, the ISGMR and ISGQR components were distinguished on the basis of MDA, leading to excitation-energy values for these resonances consistent with the known ISGMR and ISGQR energies for the nearby stable nucleus $^{58}$Ni (see Table~1).
 
This experiment, as noted, employed deuterium gas as the target, even though inelastic scattering of $\alpha$ particles had been established for a long time as the preferred method to excite the ISGMR and there was a large data set validating the MDA in extracting the ISGMR strength distributions based on DWBA calculations. Also, break-up of deuteron adds significantly to the ``background'' in the final spectra: since the detector had to be optimized for detection of very low deuteron energies, it was not possible to separate protons from deuterons based on range versus charge measurements. Indeed, the background shown in Fig. 1 of (\cite{ni56}) arises primarily from deuteron break-up and was estimated from direct kinematic measurements for $^{58}$Ni at 50 MeV/nucleon performed previously; see (\cite{dxp}). As it happens, the choice of deuterium gas in MAYA had to do with a major practical consideration in that gaseous detectors spark at high voltages when filled with pure helium gas. A further difficulty arises because of the need for a ``mask'' to absorb the electrons resulting from the high ionization of the gas by the incoming beam (\cite{mask}). These aspects are discussed further later.

Although, as an isoscalar particle, the deuteron may be thought of as a probe ideally suited for investigation of the ISGMR, it has not been used much for the purpose, save for some very early experiments carried out in France (see, for example, Refs. \cite{ddprime1,ddprime2}). Specifically, Willis et al. had performed $(d,d')$ measurements on several nuclei using a beam of 54 MeV/nucleon energy and extracted strength distributions of various giant resonances based on a peak-fitting analysis, a method deemed less reliable than the MDA technique currently in use. As deuteron appeared to be the probe of choice for measurements with radioactive ion beams, it was important to validate the results obtained in $(d,d')$ work via direct comparison with results obtained from inelastic $\alpha$ scattering.
This was carried out by Patel et al. (\cite{Darshana2013,Patel_Thesis,Patel_Book}) at RCNP. In this first investigation of giant resonances with a deuteron probe at a beam energy amenable to cross sections for excitation of ISGMR required for radioactive ion beam experiments, they employed a 196-MeV deuteron beam to obtain the now standard at RCNP ``background free'' inelastic scattering spectra for $^{116}$Sn and $^{208}$Pb. They extracted ISGMR and ISGQR strength functions using the MDA technique [see Fig. 4 of (\cite{Darshana2013})] and demonstrated that the ISGMR strength may be extracted reliably with the deuteron probe. The properties of ISGMR and ISGQR extracted in this work agree very well with the previous values from inelastic $\alpha$ scattering [see Tables 2 and 3 in (\cite{Darshana2013})], establishing, in the process, that small-angle deuteron inelastic scattering can serve as a reliable tool for investigation of ISGMR in nuclei far from stability, using inverse-kinematics reactions.

As  mentioned earlier, there are some practical issues that initially dissuaded experimentalist from using $(\alpha,\alpha')$ in inverse kinematics measurements with radioactive ion beams. A major one was the sparking of pure helium gas at high voltages typically used in the AT-TPC systems. This may be alleviated by quenching the gas with a suitable admixture; however, the concern always was that the admixed gas (CF$_4$ or CO$_2$ have been the possibilities generally considered) would give rise to unacceptably high backgrounds, especially because of much higher expected cross sections for scattering off $^{12}$C. Another possible complication arose from the high amplification of the electrons emanating from ionization of the detector gas by the beam particles, with these electrons overwhelming the collecting wires in the TPC. This problem was solved by use of an electrostatic ``mask'', placed just below the beam (\cite{mask}); this device suppresses the collection of electrons generated by the beam on the central wires. [Such a ``mask'' was not essential in the $(d,d')$ measurements because the electron amplification is significantly smaller.] 

In two measurements, performed also at GANIL, the aforementioned problems associated with using $(\alpha,\alpha')$ for ISGMR using an AT-TPC device were  overcome to some extent (\cite{ni56_2,ni68_1,ni68_2}). Helium gas was used at a pressure of 500 mbar in MAYA, with 5\% of CF$_4$ as the quencher. Two different radioactive ion beams, $^{56}$Ni ($\sim$2$\times$10$^4$ pps) and $^{68}$Ni ($\sim$4$\times$10$^4$ pps), both at 50 MeV/nucleon incident energy, were employed, with data taken on $^{68}$Ni also for $(d,d')$ to obtain a direct comparison between the two probes; the statistics in the $(d,d')$ data was much lower than in $(\alpha,\alpha')$ because of the lower cross section for $(d,d')$, as also the experimental conditions of pressure and high voltage employed in the two experiments (\cite{ni68_1,ni68_2}). In all cases, a number of narrow peaks are observed in the inelastic-scattering spectra sitting atop, so to speak, a rather large background.

In $^{68}$Ni, a ``straight'' background was subtracted from the inelastic-scattering spectra. Complementary analyses of fitting a number of peaks to the spectra, and performing MDA, led to identification of ISGMR strength over $E_x$ = 11--23 MeV, with a dominant component at $E_x$ = 21.1 MeV; this result is consistent with the $(d,d')$ data. A possible indication of a soft isoscalar monopole resonance, mixed with ISGDR strength, was also found at 12.9$\pm$1.0 MeV, in the fitting method in the $(\alpha,\alpha')$ data. 
In case of $^{56}$Ni (\cite{ni56_2}), the background was quite large and its shape was approximated by a polynomial of order 4. A total of 9 peaks were identified in the spectra and used in peak-fitting.  The centroid position of the ISGMR was found to be 19.1$\pm$0.5 MeV which compares well with the value 19.5 MeV obtained in the previous measurement on this nucleus (\cite{ni56}). The authors also reported identification of the ISGDR strength over $E_x$ =1 0--35  MeV, albeit with large uncertainties; the observed strength distribution was consistent with the predictions of the Hartree-Fock-based RPA calculations from (\cite{naftali}). 

All these measurements were plagued by low statistics, high background, low energy resolution, and limited excitation-energy range, rendering accurate and unambiguous determination of the ISGMR and ISGDR strengths very difficult and leading to large uncertainties. Nevertheless, their success, and true importance, lies in establishing the suitability of the AT-TPC set-up for measuring the properties of unstable and exotic nuclei available at rare-isotope beam facilities the world over. It is highly likely that the forthcoming new AT-TPCs, currently under development (\cite{mittig-ahn,pancin}) will provide significantly higher energy and angular resolutions, making high-quality, small-angle inelastic-scattering measurements without these problems feasible.

A measurement of ISGMR strength in the doubly-closed-shell nucleus $^{132}$Sn has been carried out recently at the RIKEN RIB Facility in Japan (\cite{ota-garg}). The newly-developed active-target system, CAT (\cite{ota_cat}), was employed with deuterium gas. The primary aim of this measurement is to obtain a more precise value for the asymmetry term, $K_{\tau}$, of the nuclear incompressibility, in conjunction with the ISGMR data available on the stable Sn isotopes (\cite{Li_2010}). [An ``experimental'' value of $K_\tau$ can help constrain the parameters of nuclear symmetry energy (\cite{Darshana2012}).] Results from this measurement are awaited. Because of the relatively high intensity of $^{132}$Sn beams available at RIKEN ($>$ 5 x 10$^4$ pps), it is anticipated that the ISGMR would be observed with better statistics than what has been the case so far in other measurements.

As a harbinger of new and exciting things to come is that a proposal for measurement of ISGMR in $^{132}$Sn has recently been approved for the first round of experiments at the new rare isotope facility FRIB, expected to start operations in Spring, 2022 (\cite{frib}). This experiment will employ the AT-TPC from the National Superconducting Cyclotron Laboratory at the Michigan State University (\cite{Bradt2017}) and use pure helium gas. This has become feasible because this instrument operates at a much lower voltage than MAYA, eliminating the problem of sparking mentioned previously. Another innovation in the present measurement is the coupling of the AT-TPC with the S800 magnetic spectrometer (\cite{Ayyad2020}). By triggering the AT-TPC by residuals from the inelastically-scattered Sn isotopes detected in the focal plane of S800, it would be possible to obtain inelastic-scattering spectra practically free of any contributions from elastic scattering, which was a bane of previous measurements (\cite{jaspreet}).

A possible, and novel, way of measuring the compression-mode giant resonances in unstable nuclei using stored beams has recently been introduced (\cite{exl_gsi}). In a demonstration experiment with a stable nucleus, carried out at GSI, Darmstadt, a $^{58}$Ni beam of 100 MeV/nucleon energy was incident on a He gas-jet target internal to ESR, the heavy-ion storage ring at GSI. Luminosities of the order of 10$^{25}$--10$^{26}$ cm$^2$ sec$^{-1}$ were achieved in the measurement, and inelastically scattered $\alpha$ recoils were measured at very forward  center-of-mass angles ($\theta_{c.m.}\leq1.5^{\circ}$) with ultra-high vacuum compatible detectors. The results indicated a dominant contribution from the ISGMR, exhausting 79$^{+12}_{-11}$\% EWSR at an excitation energy consistent with the previously reported values for ISGMR in this nucleus. This experiment used a stable beam (the luminosity for an unstable beam would have been low for obtaining meaningful results in the short-time period assigned for the experiment) and the statistics in the final spectra were rather low; still, it points to the possibility of such measurements with the advent of future facilities, such as FAIR, when much higher luminosities for unstable nuclei will become feasible.

\section{\textit{Summary and Outlook}}
In this chapter, we have provided a brief overview of the history, development, and current status of investigations of the isoscalar giant resonances, which are highly collective oscillations of atomic nuclei. Specifically, we have discussed the isoscalar giant monopole resonance (ISGMR; $L$ = 0); the isoscalar dipole resonance (ISGDR; $L$ = 1); isoscalar giant quadrupole resonance (ISGQR; $L$ = 2); and the isoscalar high-energy octupole resonance (ISHEOR; $L$ = 3). All of these have been investigated in varying detail over the past half century. In that sense, this is what one would consider a ``mature'' field of research. Still, it remains an active field of continuing work, especially on the ISGMR because of its direct connection to an ``experimental'' value of the incompressibility of nuclear matter, $K_\infty$. The latter quantity is of importance not only to nuclear structure studies, but also to the equation of state (EOS) of dense matter, the neutron stars in particular.

The work on the ISGMR is expected to get further fillip with the advent of the facilities for rare-isotope ion beams -- RIBF (Japan), FRIB (USA), FAIR (Germany), RAON (South Korea), and HIAF (China). [Of these, the RIBF has now been operating for many years and has produced, and continues to produce, many important results.] The possibility of identifying this resonance in nuclei far from the stability line can be expected to provide important information on many aspects of the nuclear EOS, especially the asymmetry term of the nuclear incompressibility, $K_\tau$.

The author wishes to most gratefully acknowledge Dr. Kevin Howard and Mr. Joe Arroyo for their invaluable assistance with preparations of some of the figures and the tables. This work has been supported in part by the U. S. National Science foundation (Grant No. PHY-2011890).

\bibliographystyle{aps-nameyear}
\bibliography{Garg_1}
\nocite{*}
\end{document}